\documentclass[sigconf]{acmart}\pdfoutput=1

\usepackage{tabularx, colortbl}
\usepackage{verbatim}
\usepackage{array}
\usepackage{booktabs}
\usepackage{longtable}
\usepackage{multirow}
\usepackage{enumitem}

\newcommand*\circled[1]{\raisebox{.5pt}{\textcircled{\raisebox{-.5pt}{#1}}}}

\def\toolname{\textsc{DocML}}

\copyrightyear{2023}
\acmYear{2023}
\setcopyright{acmlicensed}\acmConference[CHI '23]{Proceedings of the 2023 CHI Conference on Human Factors in Computing Systems}{April 23--28, 2023}{Hamburg, Germany}
\acmBooktitle{Proceedings of the 2023 CHI Conference on Human Factors in Computing Systems (CHI '23), April 23--28, 2023, Hamburg, Germany}
\acmPrice{15.00}
\acmDOI{10.1145/3544548.3581518}
\acmISBN{978-1-4503-9421-5/23/04}

\begin{document}

\title{Aspirations and Practice of ML Model Documentation: Moving the Needle with Nudging and Traceability}

\author{Avinash Bhat}
\authornote{Both authors contributed equally to this research.}
\affiliation{%
  \institution{McGill University}
  \country{Canada}
  }

\author{Austin Coursey}
\authornotemark[1]
\affiliation{%
  \institution{Vanderbilt University}
  \country{United States}  
  }

\author{Grace Hu}
\affiliation{%
  \institution{McGill University}
  \country{Canada}
  }

\author{Sixian Li}
\affiliation{%
  \institution{McGill University}
  \country{Canada}
  }

\author{Nadia Nahar}
\affiliation{%
  \institution{Carnegie Mellon University}
  \country{United States}
  }

\author{Shurui Zhou}
\affiliation{%
  \institution{University of Toronto}
  \country{Canada}
  }

\author{Christian Kästner}
\affiliation{%
  \institution{Carnegie Mellon University}\country{United States}
  }

\author{Jin L.C. Guo}
\affiliation{%
  \institution{McGill University}
  \country{Canada}
  }
  
\renewcommand{\shortauthors}{Bhat and Coursey, et al.}

\begin{abstract}

The documentation practice for machine-learned (ML) models often falls short of established practices for traditional software, which impedes model accountability and inadvertently abets inappropriate or misuse of models. Recently, model cards, a proposal for model documentation, have attracted notable attention, but their impact on the actual practice is unclear. In this work, we systematically study the model documentation in the field and investigate how to encourage more responsible and accountable documentation practice.  Our analysis of publicly available model cards reveals a substantial gap between the proposal and the practice. We then design a tool named \toolname{} aiming to (1) nudge the data scientists to comply with the model cards proposal during the model development, especially the sections related to ethics, and (2) assess and manage the documentation quality. A lab study reveals the benefit of our tool towards long-term documentation quality and accountability.
\end{abstract}



\maketitle

\begin{comment}
Austin's material google drive: https://drive.google.com/drive/folders/1NOBWp3IpR7H-zDgC68Y433rY0m5gatS9
\end{comment}

\section{Introduction}
Documentation serves as the primary resource to understand and evaluate reusable software components when adopting them in developing applications~\cite{doc_practitioners_perspective}.
Machine-learned (ML) models increasingly are integrated as components into software
systems and would benefit from similar documentation~\cite{NZLK:ICSE22, se4ml_case_study, hulten2018building}.
Stakeholders, including data scientists, AI engineers, domain experts, and software engineers, resort to the documentation to answer questions such as what use cases are supported, what performance to expect, and what ethical and safety impacts to consider once the model is deployed in applications at scale. 
Nevertheless, ML models shared as pretrained models or services are often poorly documented. Still they are reused in many applications, sometimes in applications for which they were not designed. Serious issues related to misuse of ML models have been observed in various applications, notably in face recognition and tracking~\cite{buolamwini2018gender}, recruitment~\cite{amazon_recruiting_tool}, and criminal risk assessment ~\cite{dressel2018accuracy}, leading to broader concerns about their impact on social justice. 

As a reaction to observed problems in ML model reuse and accountability, significant efforts towards documenting models~\cite{modelcard,arnold2019factsheets} and data~\cite{gebru2018datasheets} have been proposed. This line of work has attracted considerable attention -- for example, 
the paper on \textit{model cards} published at FAccT 2019~\cite{modelcard} is heavily cited, and the
popular model hosting site HuggingFace has adopted the term \textit{model card} in their
user interface and guides their users to provide documentation~\cite{huggingface-modelcarddocs}.
Yet, it is largely unknown how these proposals have impacted the practice of documenting ML models and datasets. 

In this work, we systematically study ML model documentation in the field and investigate how to encourage more responsible and accountable model documentation practice. While past work has already shown often limited documentation during model development, such as few markdown cells in public notebooks and missing README files in notebook repositories on GitHub~\cite{github2017dataset,pimentel2019large}, we focus on external documentation of reusable ML models and services.
We start by investigating how the ML models and services made available are documented, in particular, how they meet the model cards template proposed by \citet{modelcard}. Our study reveals that despite adopting the model card terminology, most model development teams fail to provide meaningful and comprehensive documentation that can support scrutiny for model adoption. Certain aspects of documentation are especially limited across different contexts of model development (i.e. open-source and proprietary), such as information regarding the data collection process, evaluation statistics explanation, and concrete ethical measurements. 

Motivated by the observed low adoption rate of the model cards proposal and frequent poor documentation quality even when the proposal is followed, we explore how we could improve the adoption of model cards and encourage good documentation practices. To this end, we design and implement a documentation tool for data scientists, named \toolname{}, that supports creating and updating user-oriented documentation during model development in the computational notebook environment. A user study with 16 participants demonstrates that, when \toolname{} was presented, data scientists adopted documentation approaches that benefit accessing and managing model documentation quality. They also showed more deliberate consideration of model development context and ethical concerns.

Our work makes the following contributions to understanding and supporting ML model documentation practice:

\begin{enumerate}[topsep=2pt]
   \item Results from our empirical study, that delineate the current practice of public model cards and highlight a clear gap between the information needed for the model users and information provided by the model developers; 
    \item A rubric for evaluating ML documentation, developed and used in our study based on model cards proposal, which can be adopted by model developers and users as a documentation guideline or quality assessment tool; 
    \item A JupyterLab extension, \toolname{} to support data scientists to write, inspect and maintain model cards during the model development process, evaluated in a user study. 
\end{enumerate}

The artifacts created in this study including the rubric, list of assessed model cards, user study design, and \toolname{} source code, are shared as supplementary materials  alongside the paper in the ACM Digital Library to support future investigation on improving ML documentation.

\section{Related Work}
In this section, we discuss how our work is situated in previous literature on software documentation, machine learning documentation, and tool support for data scientists. 

\subsection{Software Documentation}
Documentation plays a key role in various software qualities, such as usability and maintainability~\cite{Ian2016SE}. The primary information about the objectives, design, and usage of the software is recorded in different types of documentation. The study of software documentation concentrates mostly on the aspects of documentation property and quality~\cite{aghajani2019software, prana2019categorizing, arya2020information}, documentation search and discovery~\cite{stylos2006mica, STOLEE201635}, content augmentation~\cite{treude2016augmenting, robillard2017demand}, and documentation creation support~\cite{moreno2013automatic, torii, hellman2021generating}. Among them, our work is most relevant to the previous inquiry on documentation quality and interactive documentation creation support. 

% The understanding of software quality is mainly acquired from investigating how software stakeholders consider and discuss documentation related problems. 
Through a survey study with 323 software professionals at IBM, Uddin and Robillard identified ten common API documentation problems that manifested in practice~\cite{7140676}. Among those problems, \textit{incompleteness} and \textit{ambiguity} were considered the most frequent problems that caused severe impacts. A recent study by Aghajani et al. examined documentation problems through a data-driven approach~\cite{aghajani2019software}. They developed a taxonomy of documentation issues by analyzing the documentation-related discussion developer mailing lists, Stack Overflow discussions, issue repositories, and pull requests. \textit{Completeness} and \textit{up-to-dateness} are frequently mentioned. Together with \textit{correctness}, they constitute the category of issues concerning documentation content. At the same time, issues beyond documentation content are extremely common and have profound implications for the documentation writers, readers, and maintainers, such as how the content of the documentation is written and organized (e.g., documentation usability and maintenance), documentation process (e.g., traceability and contribution), and documentation tool (e.g., bugs, supports, and improper tool usage). This taxonomy illustrates the complexity of documentation concerns and calls for a consideration of documentation within the context of software development.

A large body of work on supporting documentation creation aims to automate content generation. Examples include generating progress-related documentation such as commit messages~\cite{6975661} and summarizing method~\cite{mcburney2014automatic}, files~\cite{moreno2013automatic}, or even the whole project~\cite{hellman2021generating}. Such work normally relies on heuristic or machine learning methods to extract or synthesize content from the input artifacts and inevitably introduces both errors and biases. Since our work emphasizes documentation quality, a more interactive approach with which the documentation writer has full control over the content being created is more relevant. The work by Head et al. is an example of focusing on the interaction aspect for tutorial writers~\cite{torii}. 
We adopt a similar approach. Motivated by the empirical observations of model documentation quality, we seek to address the needs from data scientists during model development and documentation through the interaction design.  

\subsection{Documentation for Machine Learning}
\sloppypar
\label{subsec:doc_ml}
The interest in ML documentation mostly concerns data and model documentation~\cite{boyd2021datasheets}, proposing what content to include in such documentation. On the data side, work on \textit{Dataset Nutrition Labels}~\cite{holland2020dataset} and \textit{Datasheets}~\cite{gebru2018datasheets} propose standards for documenting information related to the data, such as provenance, statistics, and accountable parties. Using software modeling techniques, the work of \textit{DescribeML} proposes to describe the dataset structure, provenance, and social concerns (e.g. biases, potential harm, and privacy) through a domain specific language~\cite{giner-miguelez_describeml_2017}. On the model side, \textit{model cards}~\cite{modelcard} and \textit{fact sheets}~\cite{arnold2019factsheets} propose standards for model documentation. Particularly, the work on model cards proposed by \citet{modelcard} has gathered substantial interest from both academia and industry. It aims to standardize ML model documentation; it suggests that the model cards should record information beyond performance characteristics, including intended and out-of-scope use cases, potential pitfalls, and ethical considerations. Several companies such as Google, Nvidia, and Salesforce have adopted model cards for some of their public models. Hugging Face, a open-source ML model hosting platform, also encourages its users to adopt model cards when sharing their models. Our work provides a more critical view of the current adoption of model cards. We set off to understand the impact of the model cards proposal on the quality of model documentation.

Previous work on the ML documentation process is relatively scarce.
For \textit{fact sheets}, \citet{richards2020factsheets} discuss that an interactive process with stakeholders is needed to define what information should be contained in the documentation in the first place. The proposed methodology describes how stakeholders can instrument their documentation generation at each stage of the AI development life cycle by asking concrete questions relevant to that stage. While the outcome of this process might resemble the model cards, it provides more support for planning the documentation effort and collaboration between different roles within the organization towards AI documentation. A more recent work built on the concept of model cards delves deep into how the content of model cards can support non-experts in making decisions related to the model~\cite{crisan_interactive_2022}. Compared to the standard model cards proposed by \citet{modelcard}, this work and its notion of \textit{``interactive model cards''} shift the focus to the model users and provide more probing and assessing support for them to better understand the risks and ethical consequences related to model adoption.

Despite the intense interest in ML documentation, in practice, the effort and therefore the quality of documentation still fall short. In a recent study with 45 practitioners from 28 organizations, documentation is identified as one of the biggest challenges when building and deploying ML systems into production~\cite{NZLK:ICSE22}. Similar to traditional software, incomplete and outdated documentation is a major concern. On the data side, the existing data documentation for public datasets is ``never sufficient for model teams to understand the data''~\cite{NZLK:ICSE22}. On the model side, missing documentation causes hidden assumptions and losing knowledge on key decisions about the models being developed. Our work contributes to filling the wide gap between aspiration and practice for ML documentation. In particular, as a starting point, we aim to nudge the data scientists to consider various aspects of model cards during model development and adopt better practices towards model documentation.

\subsection{Tool Support for Data Scientists}
As part of the ML development team, data scientists fulfill a decisive role in shaping the machine learning pipeline~\cite{muller2019data}. From available public or internal data sets, data scientists perform complex data wrangling to understand and transform the data into usable formats for ML models. They also experiment with different model architectures and hyper-parameter settings to improve the model performance on important metrics. The entire process is iterative and often performed on computational notebooks, such as Jupyter notebooks and Google Colab~\cite{zhang2020data}. Computational notebooks are effective when used as a scratch pad to quickly test ideas or to create a narrative computational storyline, but at the same time, they are reported to suffer from many problems, such as missing version control, unpredictable executing orders, and ill support for managing dependencies and debugging, creating significant barriers for data scientists to adopt best engineering practices and to reliably develop ML models~\cite{souti2020notebook}.

Existing work supporting data scientists' workflow mainly focuses on the notebook environment. The effort includes but is not limited to synthesizing data wrangling code~\cite{ian2022wrex}, visualizing and comparing alternative paths in the ML pipeline~\cite{wang2022diff, nathaniel2022fork}, and supporting cleaning of exploratory code~\cite{head2019managing}. In terms of documentation, \citet{yang2021subtle} proposed a documentation tool WrangleDOC to automatically summarize the data wrangling code in the notebook through program synthesis. The generated summary consists of data input, output, and transformation that can assist the data scientists in understanding and verifying the early steps of the ML pipeline. Along similar lines, \citet{wang2021documentation} proposed a tool called Themisto that combines deep-learning and information retrieval approaches to generate documentation for code cells in the notebook. Themisto also prompts the data scientists to add documentation for the code cells with output. Compared with those works, our tool has a distinct objective. The target users of WrangleDOC and Themisto are data scientists themselves. Those tools aim to support documenting the notebooks for data scientists to develop and reuse notebooks. Therefore, the resulting documentation can be filled with developmental details.
\citet{mlte} propose a domain-specific language for model evaluations that require embedding in the development process and that can automatically generate reports as documentation including many different facets of model quality.
The tool proposed in our work, however, focuses on encouraging data scientists to consider ethical aspects of their model development and to follow the documentation standard to create documentation for various stakeholders who need to reason about whether the model properties in different use cases.

\section{Understanding Model Card Practice}

The number of ML models being published and reused is increasing at an astounding speed. \textit{Hugging Face}, one of the popular platforms for sharing and hosting reusable models, is used by more than 5,000 organizations and currently hosts more than 19 thousand machine-learned models~\cite{huggingface}. The top-ranked model on Hugging Face named \textit{BERT base model (uncased)}\footnote{https://huggingface.co/bert-base-uncased} is downloaded more than 22M times per month (accessed on August 2022).
Many organizations also offer proprietary models for a wide range of tasks through public APIs, from BigTech companies such as Google and AWS to many startups.

The technical steps for reusing models and incorporating them as components into applications for various predictive tasks are easy, typically by downloading the trained model binary or calling a REST API.
However, understanding the scope and quality of a model is often not obvious.
Incomplete documentation of ML models can cause serious trouble for potential model adopters to properly set up the models within their own application. More importantly, without information about the model development process and their impact on performance and ethics in the application domain, the models might be misused or used without proper care, therefore causing various harm to the end-users~\cite{blodgett-etal-2020-language, buolamwini2018gender}.

To understand the current practice in documenting reusable models and the gap between recommendations and practices, we conduct an empirical study on model documentation.
We start with collecting a dataset of models and corresponding documentation that explicitly or implicitly indicates the adoption of the proposal of \textit{model cards}. We focus our study on model cards because this format has had a considerable impact in both academia and industry~\cite{modelcard} (more in Section~\ref{subsec:doc_ml}). We then examine the collected model cards using a rubric we created based on the original proposal for model cards~\cite{modelcard}. Our analysis is entirely manual and mixes qualitative and quantitative aspects to ensure the reliability of our evaluation. We developed and validated our rubric iteratively and release it publicly as a potential foundation for documentation guidelines or quality assessment tools. Finally, we discuss the implications of the model documentation quality evaluation result.

\subsection{Background: The Model Cards Proposal}
\label{subsec:model_card}
We first provide more details about the work of model cards. Before this work, ML models were mainly compared based on model performance, measured mostly by metrics on the whole test dataset, such as precision, recall, and f-measures. The proposal of ``model cards,'' suggests that the model comparisons should consider the ethical axes especially when the model is going to be adopted in applications that have a serious impact on people's lives. It further advocates that the model should be evaluated on the performance of subgroups divided by culture, demographic, other domain-relevant conditions, and their intersections. Overall, the model cards proposal suggests including nine sections for the model documentation to encourage more responsible and accountable practice during model development. The sections are:
\begin{itemize}
    \item \textbf{Model Details}, lists basic information about the model, such as model release date, its version, the type of the model, license, responsible parties, and how to contact them;
    \item \textbf{Intended Use}, describes the primary use cases and users that the model serves and the use cases that are out-of-scope but easily confused with or highly related to;
    \item \textbf{Factors,} records how the demographic or phenotypic groups, as well as instrumental and environmental factors, impact the model performance;
    \item \textbf{Metrics}, covers the measurement of model performance, including how those measurements are calculated;
    \item \textbf{Evaluation Data}, describes the details of the datasets used to quantitatively evaluate the model performance. It should also include the justification of dataset selection and any prepossessing procedure followed;
    \item \textbf{Training Data}, describes the details of the dataset used for training the model. When the information cannot be disclosed, it should provide basic information such as the distribution over groups;
    \item \textbf{Quantitative Analyses}, illustrates how the model performs through disaggregated evaluation with respect to each factor identified and their intersections;
    \item \textbf{Ethical Considerations}, discusses the ethical considerations taken during the model development, such as if the model uses sensitive data, the foreseen risks and how they are mitigated, etc.;
    \item \textbf{Caveats and Recommendations}, lists additional concerns that are not covered in previous sections.
\end{itemize}

To illustrate how to adopt model cards, the original proposal included two concrete model cards documenting two publicly available models: one smiling detection model~\cite{liu2015deep} and one toxicity detection model~\cite{perspectiveAPI}. Interested readers can refer to the original model cards paper by \citet{modelcard} for full details.

\subsection{Assessing Model Cards In the Field}
\subsubsection{Model Cards Collection}

\begin{table*}[t]
\caption{Model cards collection that is used in our empirical study on documentation quality.}
 \begin{tabular}{lll} 
 \toprule

Source & Subcategory & \# of Samples \\ [0.5ex]
 \midrule
\rowcolor{gray!10}      &   Top 100 Most Downloaded & 20  \\
\rowcolor{gray!10}       \multirow{-2}{*}{Hugging Face}                             &   Top 101-370 Most Downloaded & 30  \\
 & Model Card Toolkit            & 1   \\
 \multirow{-2}{*}{GitHub Model Card}                       &  ``Model Card'' in Project README        &  23  \\
\rowcolor{gray!10} Companies & -                                               & 28  \\
Baseline (Non ``Model Card'') & -                                               & 30  \\
 \midrule
\multicolumn{2}{l}{Total}                                                   & 132
\\[0.5ex] 
 \bottomrule
\end{tabular}
\label{tab:dataset}
\end{table*}

To understand how the model cards proposal is adopted in practice, we curated a dataset of model cards from three sources. We intentionally stratified our sample to cover mostly models that adopt the idea of model cards and to cover both commercial and research models. As a baseline for comparison of documentation practice, we also collected a set of model documentation that does not explicitly mention model cards. The final collection is summarized in Table~\ref{tab:dataset}. We describe the collection process from each source below.

\textbf{Hugging Face Model Cards.} We selected Hugging Face \cite{huggingface} as a source because of its large user base. They formally adopt the model cards proposal by providing documentation~\cite{huggingface-modeldocs} and training materials~\cite{huggingface-modelcarddocs} and show each model's README file under the label ``Model Card'' on the landing page for each hosted model. The content and structure of that README, however, are not checked when publishing models. We collected model cards from Hugging Face to observe how effective model card promotion can be. From its website, we collected all 370 models with more than 5,000 monthly downloads. We then randomly sampled 20 of these model cards from the top 100 monthly downloads, representing the most popular models on Hugging Face, and 30 from the rest of the 270 model cards, representing models with decent popularity.
% This research resulted in a representative sample of model cards for popular models on Hugging Face.

\textbf{GitHub Model Cards.} GitHub is a popular platform to share ML models, along with the code to train and evaluate the model. Some authors have adopted the model cards proposal for documenting the model in their README files. Therefore, we included GitHub to analyze the practices of adopting model cards for open-source ML models.
We used two search queries to identify candidate repositories. First, we used code search to identify repositories that used the Model Card Toolkit~\cite{modelcardtoolkit}, an open-source Python Model Card API with the search query \textit{``import model\_card\_toolkit''}. 
Second, we searched with the query \textit{``model card''} in the README files among all repositories.
We then manually validated all results, by removing any repositories that did not contain actual model cards, were included in the Hugging Face data, or were duplicates of another GitHub repository.
After validation, we identified only a single repository using the model card toolkit and 23 repositories recognizably adopting the model cards proposal in their README. This process yielded a relatively complete set of model cards for ML models shared on GitHub.

\textbf{Company Model Cards.} As the third source of model cards, we searched for models offered as APIs from companies (from Big Tech to startups). To this end, we relied on Google search with keywords ``model card'' and ``model card [company name],'' using company names including Nvidia, Microsoft, Google, Facebook, OpenAI, DeepMind, and Amazon.
% \todo{what companies did we search for specifically? any list}.
% Looking at model cards from these companies enables us to see what professional model cards could look like, as well as giving us a view into what companies see as important in model documentation.
We manually inspected the top
% \todo{how many; were we any systematic?}
results, discarding false positives, resulting in 28 model cards for commercial models.
While the resulting set is not exhaustive, its size is comparable to the size of documentation sets from other sources.

\textbf{Baseline (Non ``Model Cards'').}  We further included a sample of ML models hosted on GitHub that do not claim to have followed the model cards proposal, representing the common unstructured model documentation practice. To identify relevant repositories, we searched GitHub for three common and popular machine learning tasks for which models are commonly shared and nontrivial reuse questions arise (including ethical questions): object detection, sentiment analysis, and face generation. Among the identified candidate repositories, we sampled 30 (10 for each task) that meet the following criteria: have a README, release their pre-trained ML models, but do not mention model cards.

\subsubsection{Rubric Development}
\label{subsec:rubric}

\begin{table}[t]
\caption{Evaluation result for model documentation using our rubric. Each bar shows the percentage of model documentation that includes the information relating to the question. The vertical bar indicates the mean score across all data sources. The concrete rubric (question and their descriptions) is included in the supplementary material.}
\centering
    \begin{tabular}{@{}lll@{}} 
    \toprule
     Question & Mean Score &\\ [0.5ex]
     \midrule
      \textit{Model Description} & \includegraphics[width=0.8in, height=0.1in]{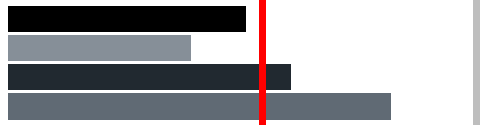} &\\
       Q1. Contact Information  & \includegraphics[width=0.8in, height=0.1in]{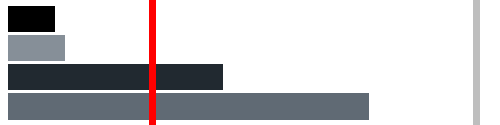} &\\ 
        Q2. Model Type & \includegraphics[width=0.8in, height=0.1in]{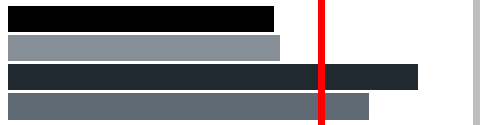} &\\
       Q3. Model Date/Version & \includegraphics[width=0.8in, height=0.1in]{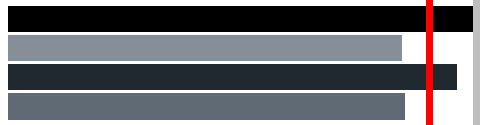} &\\
       Q4. Model License & \includegraphics[width=0.8in, height=0.1in]{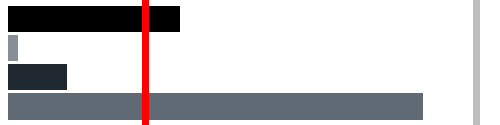} &\\ \rule{0pt}{4ex}
      \textit{Intended Usages} & \includegraphics[width=0.8in, height=0.1in]{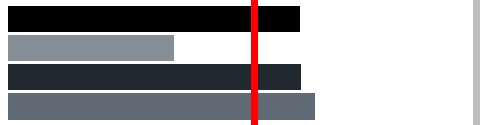} &\\
       Q5. Intended Uses & \includegraphics[width=0.8in, height=0.1in]{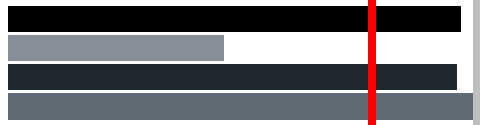} &\\
       Q6. Out of Scope Uses & \includegraphics[width=0.8in, height=0.1in]{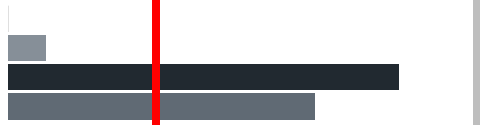} &\\
       Q7. How to Use & \includegraphics[width=0.8in, height=0.1in]{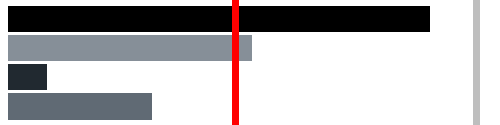} &\\
      \rule{0pt}{4ex}  
      \textit{Target Distribution} & \includegraphics[width=0.8in, height=0.1in]{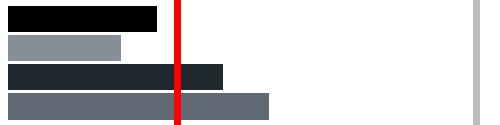} &\\
       Q8. Target Distribution Description & \includegraphics[width=0.8in, height=0.1in]{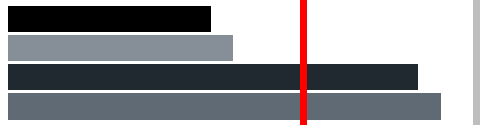} &\\
       Q9. Target Distribution Examples & \includegraphics[width=0.8in, height=0.1in]{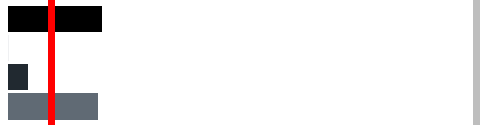} &\\ 
       \rule{0pt}{4ex}
      \textit{Evaluation Metrics} & \includegraphics[width=0.8in, height=0.1in]{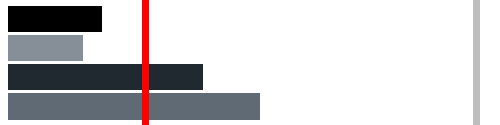} &\\
       Q10. Evaluation Statistics Reported & \includegraphics[width=0.8in, height=0.1in]{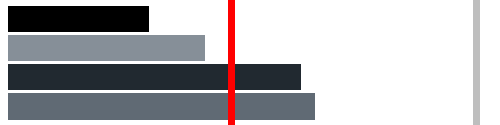} &\\
       Q11. Evaluation Statistics Explained & \includegraphics[width=0.8in, height=0.1in]{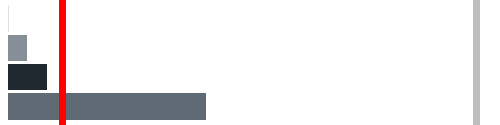} &\\
       Q12. Model Performance Visuals & \includegraphics[width=0.8in, height=0.1in]{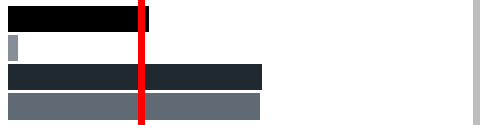} &\\ 
       \rule{0pt}{4ex}
      \textit{Evaluation Process} & \includegraphics[width=0.8in, height=0.1in]{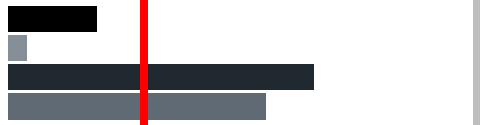} &\\
       Q13. Evaluation Process Explained & \includegraphics[width=0.8in, height=0.1in]{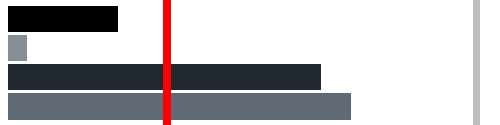} &\\
       Q14. Evaluation Data Explained & \includegraphics[width=0.8in, height=0.1in]{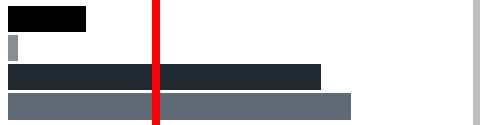} &\\
       Q15. Evaluation Data Available & \includegraphics[width=0.8in, height=0.1in]{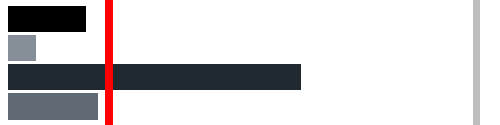} &\\ 
       \rule{0pt}{4ex}
     \textit{Training Process} & \includegraphics[width=0.8in, height=0.1in]{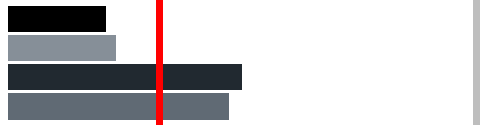} &\\
       Q16. Training Process Explained & \includegraphics[width=0.8in, height=0.1in]{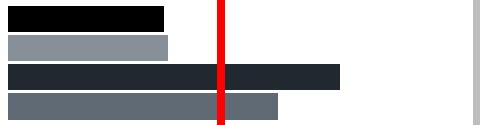} &\\
       Q17. Data Properties Explained & \includegraphics[width=0.8in, height=0.1in]{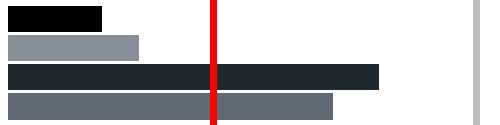} &\\
       Q18. Data Collection/Creation Explained & \includegraphics[width=0.8in, height=0.1in]{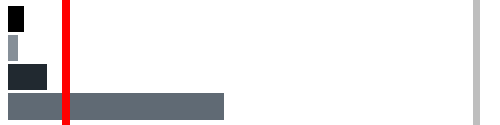} &\\
       Q19. Training Data Available & \includegraphics[width=0.8in, height=0.1in]{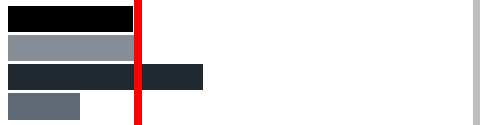} &\\ 
       \rule{0pt}{4ex}
      \textit{Ethical Considerations} & \includegraphics[width=0.8in, height=0.1in]{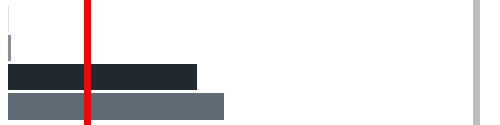} & \\
       Q20. Ethical Considerations Discussed & \includegraphics[width=0.8in, height=0.1in]{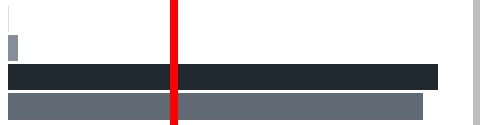} &\\
       Q21. Ethical Issue Mitigation Process & \includegraphics[width=0.8in, height=0.1in]{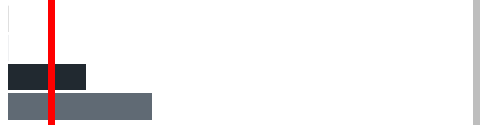} &\\ 
       Q22. Concrete Ethical Measurements & \includegraphics[width=0.8in, height=0.1in]{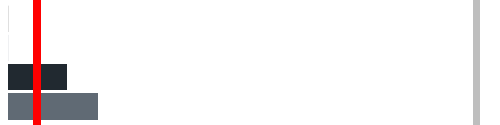} &\\
      & \includegraphics[width=0.8in]{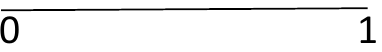}
      &\\[3.5pt]
       \multicolumn{3}{l}{$\vcenter{\hbox{\fcolorbox{white}{black}{\rule{0pt}{1pt}\rule{1pt}{0pt}}}}$ Baseline (Non ``Model Card'') $\vcenter{\hbox{\fcolorbox{white}{gray!60}{\rule{0pt}{1pt}\rule{1pt}{0pt}}}}$ Hugging Face}\\
      \multicolumn{3}{l}{$\vcenter{\hbox{\fcolorbox{white}{black!77}{\rule{0pt}{1pt}\rule{1pt}{0pt}}}}$ GitHub ``Model Card'' $\vcenter{\hbox{\fcolorbox{white}{gray!95}{\rule{0pt}{1pt}\rule{1pt}{0pt}}}}$ Company 
      $\vcenter{\hbox{\fcolorbox{white}{red}{\rule{0pt}{1pt}\rule{1pt}{0pt}}}}$ Overall Mean}\\[1ex]
     \bottomrule
    \end{tabular}
\label{tab:evaluation-results}
\end{table}

We realized in the early phases of the model cards assessment the difficulty in judging reliably how well certain aspects of a model are documented, for example, if the scope of a model is described accurately and clearly. Such judgment is highly subjective; we found low inter-rater reliability and found it challenging to define and describe levels of a measure.
Hence, we converged on an approach that measures more reliably whether certain topics are covered in the documentation with concrete yes/no questions at the cost of capturing only the presence of information in the documentation but not its comprehensiveness or correctness. The resulting list of questions served as a rubric to judge how different aspects of the model cards proposal were documented and to compare model cards from different sources.

Concretely, starting from the description of each component in the original model cards paper, we converted each aspect to be covered in the recommended model card structure into a set of concrete yes/no questions. For example, for the aspect ``primary intended usage'' in the Intended Uses section of model cards, our rubric includes a question of \textit{Does this model card (or equivalent model documentation) explain scenarios in which to use the model?} We developed those questions iteratively based on our own observations while inspecting documentation in our dataset. For example, we observed that certain aspects recommended by model cards are closely related, for example, the ``Quantitative Analyses'' and ``Ethical Considerations.'' We merged those sections and added concrete questions to resolve potential ambiguity in interpreting their categories. The resulting questions for ``Ethical Considerations'' include: \textit{Are ethical
considerations discussed?} \textit{Does the documentation discuss the used \textbf{process} for considering ethical issues with the model?} and \textit{Do the documentation provide \textbf{concrete measurements} to support the discussed ethical considerations?} (Q20 - Q22 in Table~\ref{tab:evaluation-results}). For each section, we also added the interpretation of potentially ambiguous terms in the questions, as well as examples of when to rate yes and no for corner cases.

The rubric was created iteratively and underwent three rounds of inter-rater reliability assessment. An initial rubric was designed and used by six individual raters, each evaluating the same ten model documentations (random selection of five from companies and five from Hugging Face). We then updated our rubric after investigating and resolving inconsistencies among the raters. We followed the same process during the second round on a new set of 15 model cards with an inter-rater reliability of 0.59 using Cohen’s Kappa~\cite{cohen1960kappa}. In the final round, we focused on three questions that yielded the most disagreement: questions about the target distribution of the model and the description of the training data (Q8, Q17, and Q19 in Table~\ref{tab:evaluation-results}). After clarification and refining the rubric for those questions, we reached 0.73 inter-rater reliability using Cohen's Kappa for those questions using additional 15 model cards.  

As part of the rubric, we instructed raters to only look at the primary documentation (model cards or README files), but not to follow links to papers or external data, unless the main documentation makes it clear what specific information can be found there (e.g. ``for more graphs demonstrating the model's performance, see \textit{link}'').
This was an intentional choice to focus on the core documentation that
a user might read rather than evaluate what information users could acquire by digging through academic papers or conducting their own experiments.
This aligns with the model cards proposal that suggests collecting all important information in a compact description.

After establishing the reliability of the assessment rubric, one author manually rated all model cards in our dataset using the rubric.  Note that we do not intend this rubric to give an overall score to a model card, but intend it to be used to determine whether the different aspects of information recommended by the model cards proposal are provided. The complete rubric is included in the supplementary material for future reuse and refinement by model developers and other model stakeholders.

\subsubsection{Threats to Validity}
As discussed, our rubric does not assess the completeness or correctness of information, but only whether a model card includes information related to the sections in the model cards proposal. We also largely excluded linked documents and papers from what we consider as the \textit{primary documentation}. Our results should be interpreted with these decisions in mind. Whereas we could analyze a (near) complete set of models with model cards from corporations and GitHub, we had to sample for Hugging Face and baseline (GitHub models without model cards). Our samples were not truly random; to consider the impact of the models and to keep the analysis manageable, we stratified the sample among popular Hugging Face projects and focused on models for three tasks in the baseline. Finally, our analysis was manual and relies on some subjective judgment despite our best attempts to clarify and validate the rubric.

\subsection{Assessment Results}
\label{subsec:model_doc_assessment}
\subsubsection{Quantitative Result}
\label{subsubsec:model_card_result}

In Table~\ref{tab:evaluation-results}, we summarize what aspects are included in each model card and baseline model documentation in our dataset. Below, we discuss our observation on examining the documentation across sources and aspects in the model cards proposal. We use the term \textit{model documentation} to refer to the entire documentation set we collected, including the baseline.

\textbf{Model cards provided by companies and in GitHub repositories tend to include more information corresponding to model cards proposal than model cards on Hugging Face}. Model cards provided by companies rank on top for 11 out of 22 questions based on the mean score, while Model cards in GitHub rank on top for eight questions (see Table~\ref{tab:evaluation-results}). In contrast, model cards on Hugging Face are less likely to include information related to most questions - all but related to model type (Q2), model date and version (Q3), reporting evaluation statistics (Q10), and providing training data (Q19), by Dunn's Kruskal-Wallis multiple comparisons test. We also found no significant differences between the two strata of our Hugging Face sample, suggesting that the most popular models are not documented more comprehensively than less popular ones.
Our baseline (documentation in GitHub repositories without mentioning model cards) draws a more mixed picture. They are rated similarly to or even higher than those of companies' model cards and GitHub model cards for some aspects, such as the intended use (Q5 and Q7), but fall short on questions related to ethical considerations (Q20-Q22), where they rarely include any information, similar to Hugging Face model cards.

\textbf{Most model documentation have an unbalanced coverage of aspects in the model cards proposal.} Overall, we found that 18 of the 22 types of information covered by our questions are included in less than half of the models' documentation. Only questions about the model type (Q2), model date and version (Q3), intended uses (Q5), and target distribution description (Q8) are included in more than half of the models' documentation. Q6 about the situations where a user should \textit{not} use a model, however, is only documented in 32\% of the models' documentation. Similarly, merely 35\% of the models' documentation includes some discussion of bias or ethics (Q20). The ethical issue mitigation process (Q21) and measurement (Q22) are included in less than 10\% of the models' documentation.

\subsubsection{Further Observations}
As discussed in Section~\ref{subsec:rubric}, our rubric is used to evaluate the occurrence of information in the documentation for each question, not the comprehensiveness or correctness of the information. Nevertheless, during our assessment, we did observe strong variance in the extent to which the documentation answers those questions in the rubric.  

\textbf{The information provided in the model documentation is often vague.} Taking Q8 about the \textit{target distribution} as an example: Q8 is one of the few questions that is included in more than half of the model documentation. Yet, the majority of the documentation fails to provide more than very vague or generic information about the target distribution. 
% We determined that only 28 out of the 83 model documentation provided information about target distribution (33.7\%) can be credited for including a high quality description. 
For example, one model card\footnote{\url{https://huggingface.co/GroNLP/bert-base-dutch-cased}} from Hugging Face claims to be a ``Dutch pre-trained BERT model'', hinting the target distribution being strings that represent text in Dutch -- however, it leaves many questions about other characteristics of the inputs a model user could expect for the model to work for, such as the domain of the text (e.g., news, social media, reviews, etc.) and the style of the text (e.g., verbal and written). As one of the few exceptions, Nvidia PeopleNet model card\footnote{\url{https://catalog.ngc.nvidia.com/orgs/nvidia/models/tlt_peoplenet}} represents a high-quality description of the target distribution, describing their model as being able to detect ``persons, bags and faces'' from ``RGB Image of dimensions: 960 X 544 X 3 (W x H x C).'' This model card then details the cases when the models might not perform as expected, including when target objects are smaller than 10x10 pixels, when more than 20\% of the objects are occluded or truncated, when the photo is taken in dark lighting conditions, and so forth.

\textbf{Shallow or no discussion along the ethical axes.} While the GitHub projects model cards and the model cards from companies largely include information to answer Q20 (any ethical considerations), 
% only 20 out of the 48 model cards (41.7\%) described their ethical considerations reasonably well. 
the information is often insufficient to examine concrete actual ethical issues related to the model.
For example, one model card\footnote{\url{https://help.salesforce.com/s/articleView?id=sf.mc_anb_einstein_messaging_insights_ethical_considerations.htm}} from a company simply states in their ethical considerations section: \textit{``we attempted to avoid bias and other ethical risks by not including demographic data in the model,''} but offer no further discussion, such as the scope of demographic data, the rationale of not including it, other biases from the data collection process, and steps taken for fairness auditing~\cite{10.1145/3290605.3300830}. Such narrow or shallow discussions are reflected in the drop of scores between question Q20 (any ethical discussion) and questions Q21 and Q22 that engage with the ethical issue mitigation process and concrete measurements (see Table~\ref{tab:evaluation-results}). A model card that demonstrates a more extensive explanation of ethical considerations can be seen in Salesforce's CTRL model card.\footnote{\url{https://github.com/salesforce/ctrl/blob/master/ModelCard.pdf}} It mentions that the \textit{``model was evaluated internally as well as externally by third parties, including the Partnership on AI, prior to release''} and provide a detailed description of steps they took to mitigate potential misuse, indicating that more extensive and meaningful effort towards mitigating potential ethical issues.

\textbf{Model documentation often is not self-contained and sometimes directs the readers to additional resources}. In particular, 45 out of the 132 model documentation (eight from Baseline and 37 from other model cards) in our dataset contain a link to an academic paper without summarizing the key information. In particular, 50\% of the GitHub model documentation without mentioning model cards, and 44\% of the Hugging Face model cards simply state that they are implementations or reimplementations of a specific linked academic paper. As discussed above, we do not consider this as an adequate substitute for the targeted and concise information suggested by the model cards proposal. A research paper is normally presented in a way targeting readers with sufficient academic background, not necessarily aligned with the background of the model users. Furthermore, in terms of comprehensiveness, including multiple links to outside sources decentralizes the information, increasing the chance of a model user missing key details.

% % As we discussed in Section~\ref{subsubsec:model_card_result}, these two categories consistently score worse for most of the questions in our rubric. Model creators in those cases might have provided more detailed explanations of their models in the linked papers. Even that is the case, a research paper is normally presented in a way targeting readers with sufficient academic background, not necessarily aligned with the background of the model users. Furthermore, in terms of comprehensiveness, including multiple links to outside sources decentralizes the information, increasing the chance of a model user missing key details.

% On the other hand, for the two categories with higher scores on most of the questions, i.e. Company Model Cards and GitHub Model Cards, they have demonstrated more profound understanding of the original model card work and considerable effort to improve their documentation quality. Such effort might stem from the external requirements or intrinsic motivation for improving model documentation. 

\subsection{Discussion}
Drawing from our empirical investigations on model cards in the field, we discuss the gap between the original model cards proposal and the model documentation practice as well as its implications.

\textbf{Top-down aspiration alone is insufficient to systematically improve the ML model documentation practice.} Similar to the development teams of traditional software, the development teams of ML software constantly juggle different constraints and priorities~\cite{NZLK:ICSE22}. While model documentation has been recognized as important by various stakeholders, the development team often sacrifices the effort of creating high-quality documentation for other seemingly more pressing concerns, such as a fast pace to the market. Our results also suggest that Hugging Face's attempt to promote the concept of model cards and to label the READMEs as model cards in their interface is ineffective to encourage the adoption of the model cards proposal. Indeed, on average, documentation of models on Hugging Face includes similar information to models published on GitHub without any mention of model cards. On the other hand, developers who voluntarily adopt model cards in GitHub repositories and those who publish models of companies tend to include information such as out-of-scope uses and evaluation results more systematically. Such effort might stem from the explicit requirements or intrinsic motivation for improving the model documentation.

\textbf{Certain aspects of the ML models, in particular the aspects along the ethical axes, are rarely provided in the model cards in the field.} The original model cards proposal aims to encourage model stakeholders to consider the broader context of model development and application. The ML documentation therefore should include the discussion about their decision making process during data collection and ethical consideration. However, our investigation reveals that even those who voluntarily adopt model cards commonly skip those sections. The ethical issue mitigation process (Q21) and measurement (Q22) were included in less than 10\% of the models' documentation, suggesting only a shallow (public) engagement with fairness issues. These results demonstrate how most model documentation is insufficient for reasoning about the impact of model adoption, such as the model performance on unforeseen scenarios and on minority populations.

\textbf{Meaningful encouragement for better documentation practices is needed during model development.} As with any effective software engineering practice, the practice of documentation should be placed within the context of ML model development~\cite{robillard2017demand}; it will be ill-adopted otherwise.  Hugging Face has provided the training materials for adopting the model cards proposal as part of their online tutorials. Nevertheless, our study reveals that the vast majority of its users do not take the time to follow the instructions provided; only a single one included ethical considerations or any discussion of bias. In fact, we found at least four Hugging Face model cards were created by the Hugging Face team on behalf of the model creators. Those findings indicate that ML model documentation in general still seems to be an afterthought at best. This is against the vision of the original model cards proposal that model cards should be used as an instrument to encourage more deliberate consideration of ethical aspects of ML models {\textit{during}} model development. Based on those observations, we suggest that a more meaningful encouragement for better documenting ML models should start at the model development time rather than after.

\section{\toolname{} Design}
\label{sec:interface_design}

Our work aims to accelerate wide adoption of model cards, improve compliance, and encourage more accountable documentation practice. In the meantime, we also acknowledge that ML model development is a complex process that involves many different stakeholders, such as domain experts, data scientists, and software developers. As discussed in the previous section, an effective documentation tool should fit into the concrete workflow for individuals with different backgrounds. We, therefore, focus on the data scientists in this work as a starting point, considering their critical role in shaping model development. In this section, we present an interactive documentation tool that can be integrated as part of the model development routine of data scientists. We discuss the major considerations for designing such a tool and the implementation of our prototype.

\subsection{\toolname{} User Interface}
\label{subsec:tool_design}
The primary design goal of \toolname{} is informed by our empirical analysis of model cards in the field that they are mainly ill-organized and missing important sections. \toolname{}, therefore, aims to {\textit{nudge data scientists to consider and comply with the model cards proposal during the model development, especially the sections related to ethics that are often overlooked (design goal G1)}}.

We are also inspired by the existing body of literature on ML practices, in particular the discussion related to documentation. ML models are often improved in an iterative and continuous process~\cite{NZLK:ICSE22,patel2008investigating,kery2018story,siebert2022construction}, mostly with additional data over time, there is a serious risk that model documentation and actual model properties drift apart. Since the documentation is critical for regulatory compliance, knowledge transfer, and reproducibility,  its documentation quality needs to be constantly assessed and managed~\cite{10.1007/s10664-021-09993-1, hopkins2021machine}. While computational notebooks, the tools data scientists heavily rely on, have innate support for documentation as markdown cells, they are often messy and hard to make sense of~\cite{gather2019andrewhead}. \toolname{}, therefore, aims to {\textit{support continuous assessing and managing the model documentation quality (design goal G2)}}.

Following the above two design goals, we built an early prototype of \toolname{}. We then presented the prototype to users with a data science background from our connection and solicited additional feedback. Next, we further refined the prototype to integrate those additional design considerations. Below, we describe the final interface of \toolname{} and discuss how it supports several scenarios when the data scientists develop and maintain their models and the corresponding documentation in the notebook environment.

% Difficulties: Right click cell had to be at top of cell - took a while to figure out. Didn't know how to start. Page to explain how it works would be good. Didn't know it would be at the top of the cell. Jumping you to the top would be preferred if possible.
% This vs. Checklist: Probably rather use this than a checklist. If it had worked correctly it would have done everything for him. Less work.
% How often he uses Jupyter: Haven't used it much recently. About one school year.
% Does he see a value in creating a model card: Model cards preferred. Gives a nice overview of what is going on with it. Make people more aware of what to expect.
% Workflow to encourage model card creation: Make model card tab clean (visually). Minimal effort put into it to get model card. Stay simple and not complex. 
\begin{figure*}[t]
\centering
\includegraphics[width=0.8\textwidth]{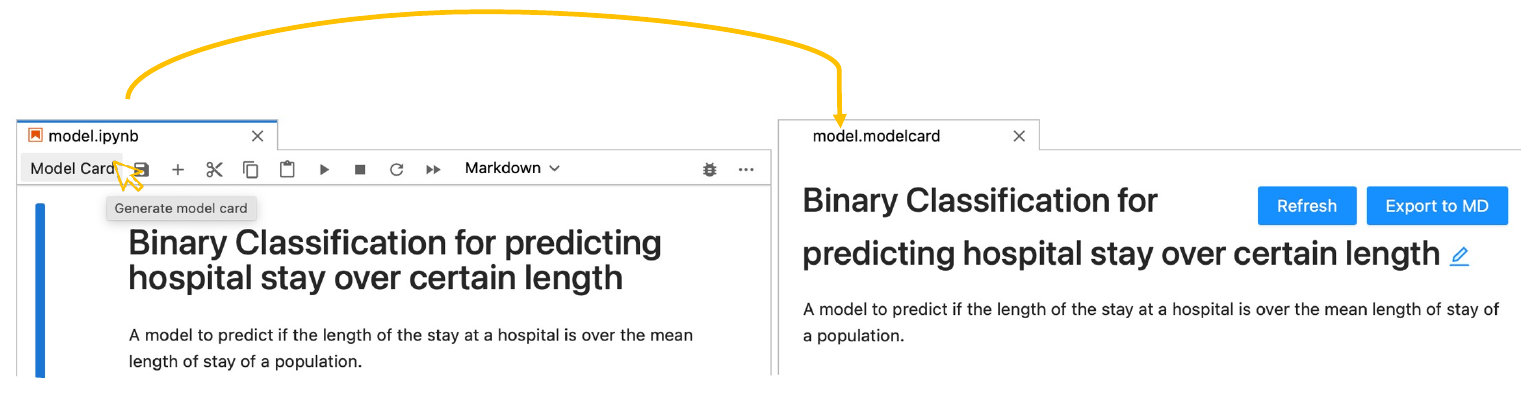}
\caption{As the data scientists are developing models, they can activate \toolname{} by clicking the tool button. The notebook and the documentation panel will show side by side on JupyterLab.}
\label{fig:side_by_side}
\Description{This figure illustrates how the users can activate DocML by clicking the tool button located on the left of tool bar. The notebook and the documentation panel will then show side by side on JupyterLab.}
\end{figure*}

\subsubsection{Creating the model cards within the notebook environment (towards G1 and G2) }
\toolname{} is designed and implemented as an extension for JupyterLab, one of the most used notebook environments for data scientists. When activated, the interactive documentation panel will be expanded alongside their notebook on JupyterLab as shown in Figure~\ref{fig:side_by_side}. The pre-defined model card sections (introduced in Section~\ref{subsec:model_card}) will show on this panel so that users will be reminded during the model development and documentation writing. Users can provide additional sections that reflecting their model development context, such as library use; they can also customize existing sections titles and orders, all through an explicit configuration file. 

\begin{figure*}[t]
\centering
\includegraphics[width=0.78\textwidth]{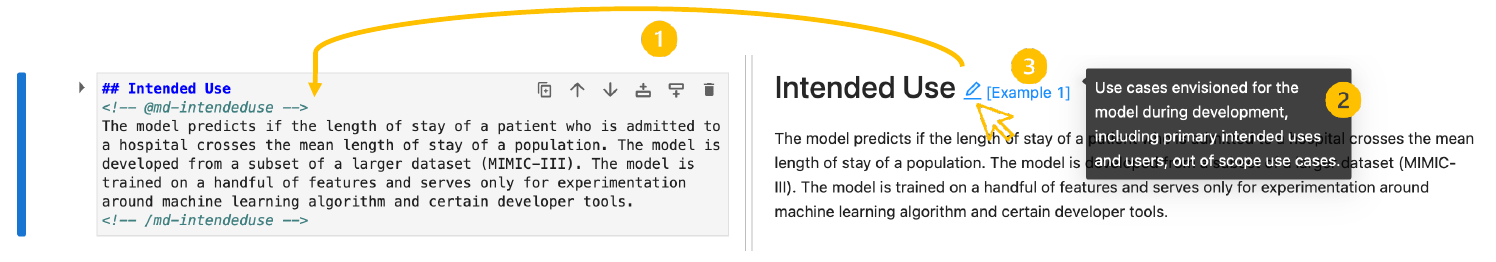}
\caption{The user starts editing the model card for each section by clicking the edit button. The content is created and maintained within the notebook \circled{1}. \toolname{} presents the description for each section when the cursor hovers over the title of the section \circled{2} and provides documentation examples through hyperlinks next to the section title \circled{3}.}.
\label{fig:md_section_desc}
\Description{This figure illustrates how the users can interact with one particular section of the model card using \toolname{}. The user starts editing the model card for each section by clicking the edit button. The content is created and maintained within the notebook. \toolname{} presents the description for each section when the cursor hovers over the title of the section and provides documentation examples through hyperlinks next to the section title.}
\end{figure*}

When users click the edit button next to the section title on the \toolname{} panel, they can start filling in the content for that section. Instead of maintaining a separate data storage for the model card, we redirect users to an automatically created markdown cell in the notebook with the section title (shown as \circled{1} in Figure~\ref{fig:md_section_desc}). To differentiate model card content that is more user oriented with other markdowns in the notebook that serve different purposes, we created a special set of HTML comments indicating their role in the model card. Users can view the latest version of model card under development anytime though clicking the \textit{Refresh} button on the panel (see Figure~\ref{fig:side_by_side}). The completed model card can be exported to a markdown file for sharing by clicking the \textit{Export to MD} button.

\subsubsection{Nudging the adoption of model cards proposal (towards G1)}
To nudge data scientists to consider and to follow the model cards proposal more effectively, we designed and implemented several features. First, when the users hover their cursor on the title of each model card section, a concise description for that section is shown to remind them what content is appropriate (as indicated by \circled{2} in Figure~\ref{fig:md_section_desc}). Moreover, explicit links to one or more examples can be added next to the section title so that users can reference when necessary (as indicated by \circled{3} in the Figure~\ref{fig:md_section_desc}). In this example, we select several high-quality model cards from our empirical study discussed in Section~\ref{subsec:model_doc_assessment} as exemplars. The section titles, their descriptions, and examples links are also customizable through the configuration file.

\begin{figure}[t]
\centering
\includegraphics[width=0.45\textwidth]{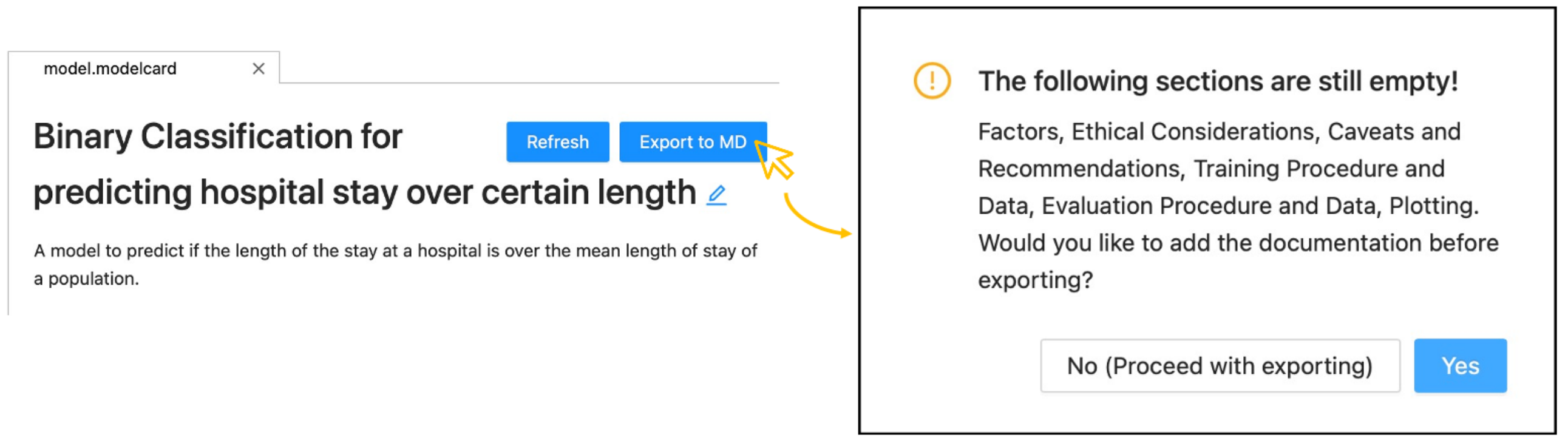}
\caption{\toolname{} suggests the users to add content in the empty sections before exporting.}
\label{fig:complete_check}
\Description{This figure illustrates how \toolname{} suggests the users add content in the empty sections before exporting.}
\end{figure}

Once the users finish editing the model card, they can export it to a markdown file for sharing. At this point, \toolname{} will perform a lightweight completion check and inform them if any pre-defined sections are still empty (Figure~\ref{fig:complete_check}). This extra step serves as an encouragement for them to complete missing sections.

\subsubsection{Model card maintenance through code-document traceability (towards G2)}
\begin{figure*}[t]
\centering
\includegraphics[width=0.9\textwidth]{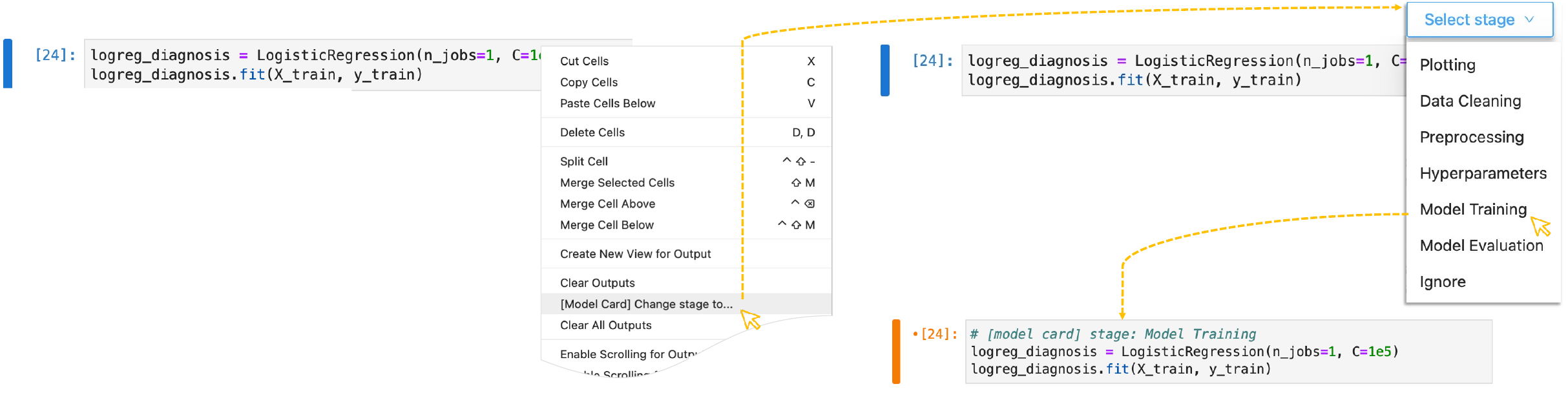}
\caption{Users can explicitly select the machine learning stages corresponding to the model documentation. Once selected, the stage is indicated through a code comment.}
\label{fig:code_stage_selection}
\Description{This figure illustrates how the users can explicitly select the machine learning stages corresponding to the model documentation using \toolname{} and how the selected stage is indicated through a code comment on the original notebook.}
\end{figure*}

Certain sections in the model card directly describe the purpose and outcome of the code cells in the notebook, such as the training process and evaluation process. The quality of those sections in this case depends on how accurate the code cells are described. To support the cross-reference between model card and source code, we adopt the concept of traceability, which is often used in software engineering for safety-critical systems~\cite{cleland2012software}. In particular, users can explicitly link the code cells to the corresponding sections in the model cards so that the content can be easily referenced and analyzed during model card creation and maintenance. We defined six stages that represent a common machine learning pipeline related to the data scientists~\cite{se4ml_case_study}, i.e. data cleaning, preprocessing, hyperparameter tuning, model training, and model evaluation. To alleviate the manual effort required on selecting the stages, we automatically identify the stages for common libraries used by data scientists, including scikit-learn,\footnote{https://scikit-learn.org/stable/} numpy,\footnote{https://numpy.org} pandas,\footnote{https://pandas.pydata.org} and matplotlib\footnote{https://matplotlib.org} through constructing a knowledge base of the API usage. Further support to other libraries can be added by enhancing the knowledge base. In case of incorrectly identified or missed code cells due to the auto-detection process, users can correct the stages manually. Under the hood, the trace links are maintained through the metadata for the code cells which cannot be easily viewed in the notebook environment. To make the information easily accessible, we automatically inject code comments to indicate those trace links, as shown in Figure~\ref{fig:code_stage_selection}. 

\begin{figure*}[t]
\centering
\includegraphics[width=0.9\textwidth]{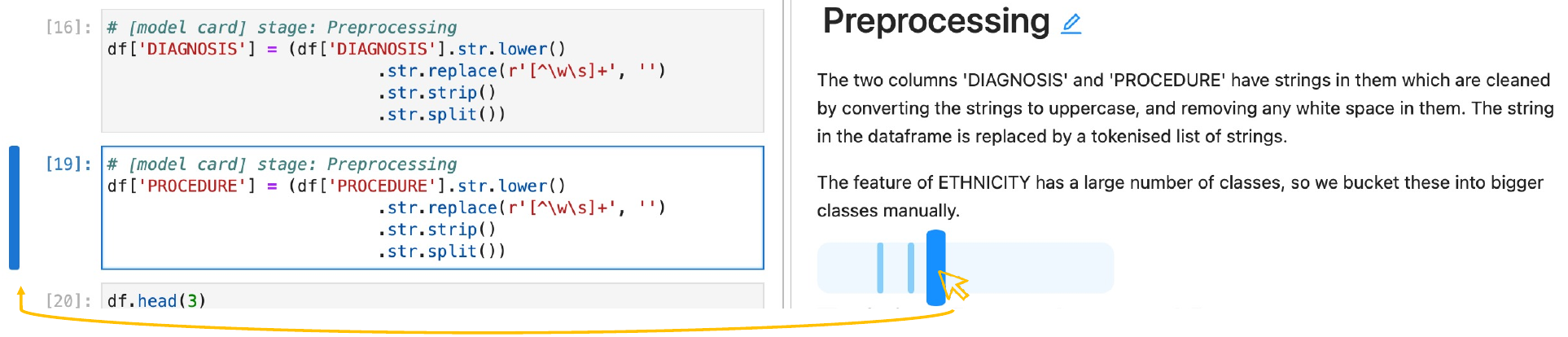}
\caption{To support the users to navigate the corresponding cells in the notebook, \toolname{} provide a navigation bar for each machine learning stage specified in the notebook.}
\label{fig:navigation_link}
\Description{This figure illustrates the navigation bar for each machine learning stage in \toolname{} that supports the users to navigate the corresponding cells in the notebook.}
\end{figure*}

Once the trace links are established, the model card maintainers or reviewers can easily navigate the code cells related to corresponding model card sections through a navigation bar on the \toolname{} panel. It also indicates the relevant location of the code cell within the notebook (see Figure~\ref{fig:navigation_link}). 

\subsection{Implementation}
% The JupyterLab environment has a server client architecture, and uses Tornado Web Server\footnote{https://www.tornadoweb.org/en/stable/} for serving the HTTP requests that are generated from the JupyterLab user interface. We leverage this decoupled architecture to create two modules: 
The architecture of \toolname{} includes a back-end extractor module, which is written in Python and JavaScript, and a front-end JupyterLab plugin written in React. It supports multiple model cards interfaces at the same time with the single back-end. The front-end module follows the interface design specifications described in Section~\ref{subsec:tool_design}. The back-end module analyzes the code in the notebook building on existing tools for program analysis,\footnote{https://github.com/andrewhead/python-program-analysis} ML stages analysis for notebook\footnote{https://github.com/yjiang2cmu/Jupyter-Notebook-Project} and cell dependencies analysis.\footnote{https://github.com/jerry-lu/cell-dependencies} Once a request is made from the front-end plugin, the content of the notebook is sent to the back-end for obtaining the cell dependencies and clustering the cells that belong to the same stage. 
%  The code cells in the notebook are first aggregated to create a single python program while maintaining the original cell indices. The program and the cell numbers are given as input to the ML stages analysis tool, which internally uses the cell dependency analysis tool to find the dependencies between the cells in the notebook. The cell dependency analysis tool depends on program analysis to convert the program into its abstract syntax tree (AST) form which is then transformed to the def-use chain. Once the cell dependencies are obtained, the ML stages analysis tool clusters the cells that belong to the same stage. 
Mappings to the relevant stage name are then added either based on the manual input by the user or based on knowledge base matching rules which classify various scikit-learn, numpy, pandas and matplotlib function calls in the sections.

The configuration file consists of a list of JSON objects, which record customizable content such as section names, their description, and any examples that showcase the suggested way of documentation. The JSON objects are populated and displayed in the front-end when \toolname{} is initialized. All markdown cells from the Jupyter Notebook are parsed to retrieve any model documentation with the specific HTML tags for displaying on the tool panel.

\section{User Study}
% Research questions:
% 1. Does our tool encourage creating more comprehensive documentation? 
% Task: TASK 1 in the user study.
% Measure: Quality of the left out sections in the Model Card.  User ratings on the model card nudging features.  

% 2. To what extent can our tool support understanding and evaluating the accuracy of the documentation and the code.
% Task: Add some inconsistency between the code and documentation and ask the participants to find them. TASK 1 in the user study.
% Measure: How much inconsistency is identified and corrected. User ratings on related features.

% 3. Can our tool prevent stale documentation during the model maintenance? 
% Task: Ask the participants to change part of the code and update related documents. TASK 2 in the user study.
% Measure: The accuracy of the documentation from TASK 2. Feedback from the participants. User ratings on related features.

We conducted a lab user study to investigate to what degree \toolname{} can support data scientists towards responsible and accountable documentation practice during model development and maintenance.

\subsection{Study design}
\label{subsec:user_study_design}
Ideally, we would have liked to observe how model developers use \toolname{} in practice over an extended period to study the subtle effects of nudging and traceability. With the low adoption of model cards and few alternatives, we found such a design infeasible. Instead, we intentionally limited the scope of our study to questions we could ask within a controlled experiment involving experienced notebook users. 
In a lab setting, we cannot well study nudging effects on model cards, since we just freshly remind all participants (both in experimental and control group) about the model cards proposal. Hence we focus our study on exploring how \toolname{} changes the focus and actions of its users, compared to users who are familiar with model cards but do not have dedicated tool support.

In a nutshell, we design a between-subject controlled experiment~\cite{blandford2008controlled}, where participants are given two notebooks with an incomplete, low-quality model card each and asked to perform tasks, including (a) selecting a model and (b) changing a model and its corresponding documentation. In the process, we observe how they navigate the notebooks, use the model cards, and make updates to the documentation. This allows us to focus on design aspects of our tool regarding tooling integration and transparency to answer the research questions below, but it provides only limited insight into the effectiveness of nudging in a natural environment -- which we leave for future studies.

Our study aims to answer three research questions: 
\begin{itemize}
    \item 
\textbf{RQ1:} What kind of information does \toolname{} encourage the data scientists to consider for ML model documentation? % (towards \textit{G1} defined in Section~\ref{subsec:tool_design})?  
\item \textbf{RQ2:} What documentation approaches emerge when
 \toolname{} is presented compared with the existing notebook environment? % (towards \textit{G2} defined in Section~\ref{subsec:tool_design})?  
\item \textbf{RQ3:} What features could be changed or added to support data scientists in model documentation? 
\end{itemize}

RQ1 relates to problems in current practices of using model cards (such as providing very selective documentation) and to our design goal $G1$ of encouraging model developers to comply more with the model cards proposal, especially regarding ethics. In contrast, RQ2 relates primarily to our design goal $G2$ of encouraging developers towards a process of continuous assessing and managing the model documentation. Finally, RQ3 seeks feedback on the tooling itself.

\subsection{Method}
We performed between-subject controlled experiments with 16 participants with or without \toolname{} in a remote lab setting using Microsoft Teams. The study protocol was approved by the research ethics board at McGill University. Below, we discuss the details of the recruitment process, lab experiments design, and how we analyzed the experiments result.

\subsubsection{Recruitment and Participants}
We recruited participants through invitations on Twitter, LinkedIn, and from our personal networks followed by screening the candidates with appropriate backgrounds. The selection criteria included having sufficient experience with the notebook environment and having shared at least one ML model prior to the study. Suitable candidates were invited to participate in the study. In the end, 16 participants were recruited. The recruited participants were asked to fill out pre-study survey form, following which they participated in the remote lab study that lasts for around one hour, and then in a post-study interview. 

Among them, 14 participants have at least one year of experience using computational notebooks. All participants have used Jupyter Notebook in academic settings (e.g. for assignments) and five of them also used Jupyter Notebook in professional settings (e.g. professional data scientists). The participants have a varied degree of experience with model documentation. While they have all developed and shared ML models with others before (the requirement for participating our study), four participants mentioned that they have not written any explicit user-oriented documentation before.  At the same time, one participant suggested that they have documented every model that they have developed.

\subsubsection{Study Process}
The 16 participants were randomly divided into two groups so that we can understand the impact introduced by our tool (RQ1 and RQ2). One group performed the study with \toolname{} (the experimental condition; participants $P_E1$-$P_E8$), and the other group without our tool (control condition; participants $P_C1$-$P_C8$). In the pre-study survey, all participants answered questions about their data scientist background and documentation practice and were informed of the model cards proposal. Participants in the experimental group were additionally asked to watch a short tutorial video introducing the functions of \toolname{} and to access and get familiar with \toolname{}'s interface. 

% Due to the restrictions on user studies during the pandemic, the lab study was performed remotely through Microsoft TEAM.
% using Microsoft Azure virtual machines. The chosen virtual machines were of Standard\_D2s\_v3 size, with 8GB memory and 2 vCPUs. 
% Three such virtual machines with identical environment were procured; one for the control group, another for the experimental group and the last one was used for the playground. 
At the start of the study, we provided the participants with an incomplete notebook and a model card in the form of a README.md file. This model card suffers from several common documentation quality issues. For example, it only contains a small number of model card sections, such as information about the model, intended use, and preprocessing. The information in each section is not necessarily complete. Moreover, the content of the model card is inconsistent with the original notebook in three places, specifically the library use, hyper-parameters, and the dataset feature description. We deliberately chose not to inform the participants of the concrete quality problems to mimic the model cards they might encounter in practice. All the study artifacts are included in the supplementary materials.

The participants from both groups were asked to perform two identical tasks, around 20 minutes each, representing common activities during ML model development and maintenance. Task 1 was to choose one among the two potential models we provided in the notebook and complete the documentation for the model of their choice. The participants were encouraged to make any changes to the existing code and documentation to improve their accuracy, completeness, or other quality attributes. During Task 2, the participants were asked to develop a new model on the same dataset using different features and to update the documentation accordingly. The resulting documentation from two groups was compared to answer RQ1. The entire process was video recorded for later analysis on their documentation activities to answer our RQ2. 
 
Upon completion of the two tasks, we interviewed the participants about their experiences related to the model documentation (RQ3). For the experimental group, we asked the participants to evaluate six major features of \toolname{} and how the features might fit in the workflow of data scientists. For the control group, we sought their opinion on the potential support that would improve their documentation experience.

\subsubsection{Analysis}
We analyze the experiment primarily qualitatively to understand how the participants approach documentation under different conditions. We first assessed the kind of changes they made to the provided model cards to answer RQ1. The analysis for RQ2 was done through a thematic analysis~\cite{Vaismoradi2013} of the video recordings by two of the authors. We particularly focused on the various activities they performed to understand data scientists' attitudes towards documentation creation and quality assessment and if they leverage the features of \toolname{} when available. 
   
\subsection{Threat to Validity}
The study is limited by the lab setting where the time constraint plays a major factor impacting the documentation experience of the participants. Task complexity, target domain, and other factors might play a bigger role in practice. Moreover, despite providing the tutorial and tool access prior to the study, participants in the experimental group still experience a learning curve of using the tool. Therefore, our study might not reflect the documentation quality developed by users who are already familiar with the interface. Finally, the long-term impact of deploying the tool, especially during continuous model and document evolution, cannot be observed from current study design.

\subsection{Results and Observations}
\subsubsection{Consideration of the Model Cards Proposal (RQ1)}
\textbf{Observation of the control group.} Most participants from the control group made small edits on one or more sections in the provided model cards. The sections that were edited most include the \textit{Hyperparameters} ($P_C1$, $P_C2$, $P_C3$, $P_C6$, $P_C7$) and \textit{Preprocessing} ($P_C3$, $P_C5$, $P_C6$). Only $P_C5$ made changes on the section of \textit{Intended Use}. Regarding the inconsistencies in the provided model card, $P_C1$ and $P_C6$ each fixed two places during the study, while the remaining six participants from the control group each fixed one. 

In terms of the new content added, the participants mostly focused on different aspects of the model evaluation, including the performance metrics, evaluation process, and evaluation data. Some participants used more descriptive section titles such as ``training procedure'' ($P_C8$) and ``test strategy'' ($P_C5$) while other times participants used generic terms such as ``model'' ($P_C1$, $P_C2$, $P_C3$) and ``result'' ($P_C4$, $P_C6$).

% Figure~\ref{fig:user_performance_rubric} demonstrate the evaluation of their final model cards using our rubrics. Here, we only show the sections added by the participants. 

% \begin{figure}[t]
%     \centering
%     \includegraphics[width=0.95\textwidth]{figures/user_performance_rubric.pdf}
%     \caption{Model card created by the participants evaluated by our rubric. Questions related to the provided content during this study are excluded from the comparison (Q2-Q6, Q16, Q17, Q19).}
%     \label{fig:user_performance_rubric}
% \end{figure}

\textbf{Observation of the experimental group.} 
The section edited most by the participants is 
\textit{Data Cleaning} ($P_E2$, $P_E3$, $P_E6$, $P_E8$). Regarding the inconsistencies in the provided model card, $P_E8$ fixed two places during the study and the remaining seven participants from the experimental group each fixed one.

In terms of the new content, the sections to which the participants added the most are  \textit{Training Procedure and Data} ($P_E2$, $P_E5$, $P_E7$) and \textit{Ethical Considerations} ($P_E2$, $P_E4$, $P_E8$). Notably, $P_E2$ added information about ethics including the sensitive features used by the model and the impact of the model if it is deployed to two specific sub-populations. Similarly, $P_E4$ and $P_E8$ added the use of race and ethnicity features in the model training. In comparison, no participants from the control group discussed ethics of the model development.

Despite providing the model cards template, participants sometimes still chose to add self-defined sections. For example, $P_E1$ added the \textit{RandomForestClassifier} to describe the model type and features used. $P_E3$ and $P_E5$ added a section called \textit{Exploratory Data Analysis} or \textit{EDA} to document the data distribution. Both $P_E4$ and $P_E6$ added a \textit{Problem Statement} in the markdown cells of the provided notebook describing the context of the model development in the model card. Occasionally, the newly added sections were named using ambiguous terms such as \textit{Results} ($P_E3$) and \textit{Refined models} ($P_E5$).

\begin{center}
 \fcolorbox{black}{gray!05}{%
  \begin{minipage}{0.99\linewidth}
    When \toolname{} was presented, the participants considered the scope of model documentation more broadly, as the model cards proposal suggested. Using the traditional notebook environment, the content of the model card was more performance-centric. In comparison, using \toolname{} under time constraints, the participants tended to add fewer new sections to the model cards. However, the sections they added are more likely about the context of the model development, including the \textit{ethical considerations} and \textit{problem context} -- the information often overlooked in public model documentation. 
\end{minipage}}   
\end{center}

\subsubsection{Documentation Approaches (RQ2)}
%  To answer: whether using \toolname{} can encourage creating more comprehensive documentation

Participants from both groups generally started their tasks by glancing through the notebooks to get familiar with their structure and content. How they approached the documentation diverged depending on whether \toolname{} was present. We identified three themes characterizing their documentation approaches when completing the model development and maintenance tasks. We describe them below and compare the differences when \toolname{} was presented versus not.

\textbf{Comparing and choosing different set of documentation.} The notebook has innate support for documentation in the markdown cells. While the model card has a distinct purpose of presenting information about the ML model, its content inevitably is closely related to some of the markdown cells of the corresponding notebook. 
% The experimental group was able to maintain consistency in the documentation between the notebook and the model card since \toolname{} ensures that the documentation added into the notebook is preserved in the model card. In contrast, the participants in the control group either added the documentation separately to the notebook or the model card, leading inconsistent information, or in some cases devoted additional effort on copying the documentation in the notebook over to the model card. 
In Task 1, which was closer to a model card quality assessment setting, the participants needed to carefully compare and ensure the correctness of the information provided in the notebook and the model card. Participants from the control group, therefore, spent considerable time comparing the notebook markdown cells and provided model cards. If they spotted any problems, they had to choose \textit{where} to fix them. While occasionally they made changes in one set of documentation and copied the content over to the other, most of the time they simply changed the model card but left the markdown cells unchanged, leading to inconsistent information (all except $P_C3$ and $P_C6$). In contrast, the problem of inconsistency between the two sets of documentation was naturally eliminated in the experimental group when they used \toolname{}, since \toolname{} ensured that the documentation for model cards would always be added to the notebook and updated in the model card view (all except $P_E5$ and  $P_E8$).\footnote{Due to a technical issue during the study, \toolname{} only became accessible to $P_E8$ at the end of Task 2.} In Task 2, which mimicked a model card development setting, we observe similar behaviors for the participants in the control group. They either added the documentation to the notebook and but not the model card ($P_C1$, $P_C4$, $P_C8$), or first added the content into the markdown file and then copied it to the model card ($P_C2$, $P_C3$, $P_C4$, $P_C5$). $P_C4$ even copied some code snippets into the model card. On the other hand, most participants in the experimental group (all except $P_E7$ and $P_E8$) ensured that the same documentation that was added to the notebook was also present in the model card by using \toolname{}.

\textbf{Locating corresponding source code.} Some of the model card sections directly describe the source code in the notebook and its outcome. The code can be scattered into multiple cells that are disconnected. We observe that the participants in the control group spent a significant amount of effort on locating the code cells when inspecting the provided model cards during the maintenance task. $P_C7$, for example, had a hard time finding the code cells corresponding to the sections related to dataset cleaning, pre-processing, and hyperparameters, and had to scroll the entire notebook several times before starting to settle on some of the code cells. $P_C1$, $P_C2$, $P_C3$, $P_C4$ and $P_C5$ showed a similar struggle. On the other hand, all participants except $P_E8$ from the experimental group used the \toolname{}'s navigation support enabled by the code-documentation trace links to help them examine the model card content. The participants either heavily relied on the navigation bar from the start ($P_E7$, $P_E9$) or increased their usage of the bar for finding the code cells as they got more used to the tool ($P_E1$, $P_E3$, $P_E4$, $P_E5$).

\textbf{Devoting attention during documentation.}
In the control group, participants paid most of their attention to examining the existing notebook and the provided model card during Task 1. Almost all participants in this group (except $P_C8$) devoted their effort to adding the documentation related to the algorithmic aspect of the model and its training or testing processes. When \toolname{} was presented, all participants from the experimental group except $P_E1$ and $P_E3$ read the prompt descriptions of model card sections to help their understanding, in particular, the sections of \textit{Factors}, \textit{Fairness considerations}, and \textit{Caveat and Recommendation}. Some of them further clicked the example links of model cards. The difference in attention between the two groups explains the observed differences in the resulting model cards in RQ1 -- the model cards from the control group heavily emphasized the model evaluation whereas the model cards from the experimental group focused more on the model development context and ethical considerations.

Additionally, \toolname{} seems to noticeably influence the participants to consider the trace links as an innate component of the documentation both during model maintenance and development. Some of them ($P_E5$, $P_E6$, $P_E7$) made edits to the trace links by modifying the existing links from code cells and/or adding new links for Task 1. For Task 2 during which they were asked to develop their own models, $P_E6$ and $P_E7$ devoted considerable effort to creating trace links for their newly added code cells. This additional effort might be motivated by the experienced benefit of code-documentation trace links in Task 1.

\begin{center}
 \fcolorbox{black}{gray!05}{%
  \begin{minipage}{0.99\linewidth}
    \toolname{} considerably alleviates the effort of assessing and managing the model documentation quality and prompts the participants towards more accountable documentation practice. In contrast to the control group, most participants from the experimental group chose to put more documentation effort into activities that are missing from the current practice, including understanding the model cards proposal (especially ethical-related considerations) and creating and maintaining doc-code trace links. Those activities can potentially bring non-negligible benefits to model documentation in the long term.  
\end{minipage}}   
\end{center}

\begin{figure*}[t]
    \centering
    \includegraphics[width=0.88\textwidth]{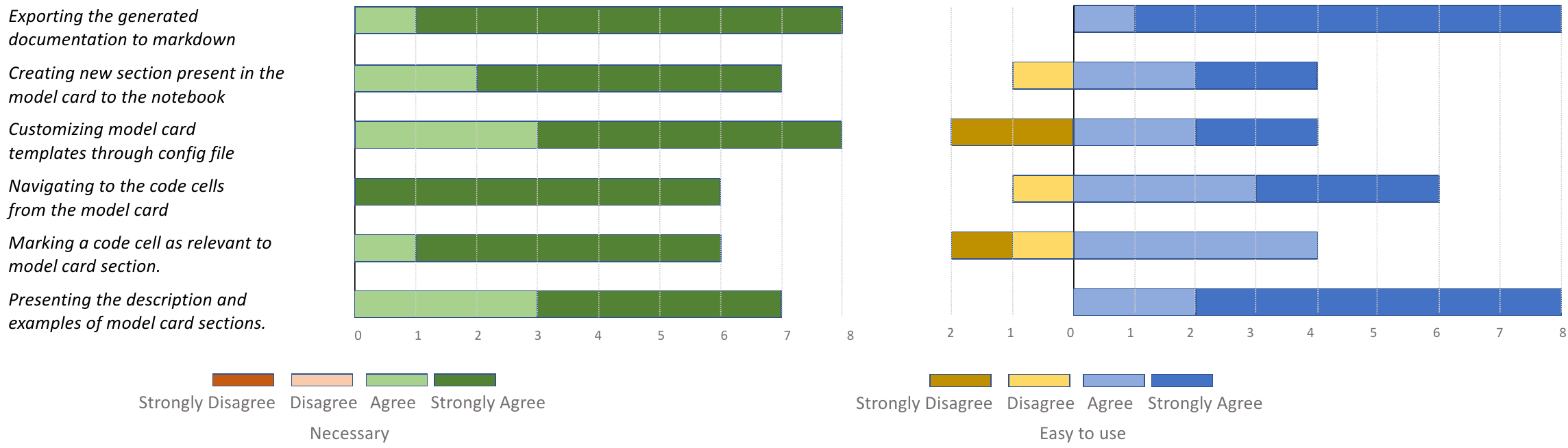}
    \caption{User evaluation on the necessity and ease of use for \toolname{}. Both aspects were evaluated through a Likert scale. Neutral and abstained input is omitted from the figure. X axis is the number of participants from the experimental group.}
    \label{fig:user_evaluation}
    \Description{This figure illustrates the user evaluation on the necessity and ease of use for \toolname{} using a Likert scale. Most participants agreed or strongly agreed that the features provided by \toolname{} are both necessary and easy to use.}
\end{figure*}

\subsubsection{User Evaluation and Feedback (RQ3)}

\textbf{Feedback from the experimental group.} Figure~\ref{fig:user_evaluation} summarizes how the participants from the experimental group rated the necessity and ease of use for different features in \toolname{}. 

Most participants agreed or strongly agreed that the features provided by \toolname{} are necessary. All participants except $P_E7$, in particular, commented on the importance of having the model card template to structure their documentation \textbf{during} the model development. They thought the prompted descriptions were helpful to better understand the model card sections. Moreover, as suggested by $P_E1$, $P_E2$, and $P_E3$, \toolname{} offers a preview of the model card during development time that is effectively separated from the notebook markdown cells. It enables the participants to consider how the information is communicated to the users. At the same time, $P_E5$ appreciated how \toolname{} can prevent inconsistency between the developmental documentation and model cards.

Most participants (all participants expect $P_E5$ and $P_E8$) also thought the navigation function supported by the trace links was especially effective so that they \textit{``do not have to see hundreds of lines of code for going to the [model card] section''} ($P_E2$). On the other hand, since the current construction of trace links requires several mouse clicks to find the stages in the model development pipeline (see Figure~\ref{fig:code_stage_selection}), $P_E1$ and $P_E3$ suggested having more intuitive options or automated solutions to achieve similar functions.

At the same time, participants from the experimental group expected more control over the section order and title in the model card. They preferred direct modification on the model card template through the UI panel rather than the configuration file, indicating a tension between customization and standardization that needs to be carefully balanced in practice. Moreover, the participants hoped the markdown cells representing the model card content could be injected next to the corresponding code cell. \toolname{} currently is limited by depending on the users to appropriately locate the newly added markdown in the notebook. 

\textbf{Feedback from the control group.} When asked what features to expect for a model documentation tool in the notebook environment, participants from the control group voiced their needs for many similar functions provided by \toolname{}, such as the documentation template ($P_C2$, $P_C3$, $P_C8$), documentation extraction from the markdown cells or code cells ($P_C1$, $P_C4$, $P_C7$), and code-documentation links ($P_C3$, $P_C6$). Those suggestions stem from both their experience during our user study and their previous experience as data scientists (and software developers). $P_C3$ and $P_C4$ specifically compared the documentation support for data scientists with more traditional software developers and pointed out that more mature tools and frameworks such as JavaDoc~\cite{JavaDoc} and Sphinx~\cite{sphinx} are unavailable. Such lack of support has caused them the most frustration during documenting models. 
% In addition, $P_C5$ also suggested a more advanced search feature on documentation based on markdown syntax (since the provided model card is in markdown syntax).

\begin{center}
 \fcolorbox{black}{gray!05}{%
  \begin{minipage}{0.99\linewidth}
    Participants from the experimental group judged the functions of \toolname{} were both necessary and generally easy to use.  At the same time, they expected improvement in flexibility and more intuitive switches between the notebook and the model card panels. 
\end{minipage}}   
\end{center}

% Participants from the controlled group voiced their needs for similar functions provided by \toolname{}, such as the documentation template and code-documentation links. Moreover, participants from both groups also suggested their preference for collaboration functions for the documentation tool. Future work is needed to understand how to fit documentation tools into the development toolchain for different stakeholders and to fit into the team dynamic. 

\section{Discussion}
In this work, we systematically investigated how the model cards proposal has been adopted in the field, finding a substantial gap between the ambitions behind model cards and actual practice, finding that model cards are rare and often shallow. Motivated by this gap and guided by literature on ML practices, we designed \toolname{} to provide data scientists direct documentation support to follow the model cards proposal and other best practices during model development and maintenance. Here, we discuss the most important findings of our work, their broad implications and limitations.

\subsection{Aspirations vs. Practice}
In all of GitHub with millions of public notebooks~\cite{pimentel2019large,github2017dataset,psallidas2019data} and many repositories sharing learning code and learned models, we found only 24 models documented explicitly with model cards. Our best effort in finding model cards published by companies also resulted in only 28 models. Considering that the model card paper is one of the most cited works on ML documentation and is often recommended, such a limited adoption indicates reluctance or difficulty to transform recommendations into standard practice.

Furthermore, even when model cards were adopted as a concept and term, they were often of low quality and provided only selective information. During our assessment, we observed strong variance in what information was provided by the model cards. Moreover, the extent to which the documentation answers those questions in the rubric also varies drastically. For example, the majority of the model cards we examined failed to provide more than vague or generic information related to target distribution and ethical considerations. Model cards found in practice were often not self-contained and sometimes directed the readers to additional resources such as research papers, thus not fulfilling the intention of providing a concise place for essential documentation. Even when digging into the papers, we often failed to find information related to the various recommended model card sections. In general, model documentation in practice seems to still be an afterthought at best. 

The original model cards paper lists several important roles that model cards serve, including improving model understanding,  helping \textit{``to standardize decision making processes for invested stakeholders,''} and encouraging \textit{``forward-looking model analysis techniques''}~\cite{modelcard}. We argue that \textbf{the current state of model documentation with model cards fulfills none of these roles well.} Previous work on interactive model cards~\cite{crisan_interactive_2022} provides an example of adopting model cards to improve the model understanding, whereas our tool \toolname{} attempts to encourage better adoption to support the other two roles of the model cards. Further studies are needed in understanding and assisting the multifaceted purpose of model cards in practice and ML documentation in general.

\subsection{Design to Facilitate Documentation Activities}
The design of documentation tools should build on the consideration of the unique characteristics of ML documentation, including how it relates to source code development and its important quality attributes. The existing \textit{Model Card Toolkit}~\cite{modelcardtoolkit} was proposed in 2020 but has barely received any adoption on GitHub. We conjecture that this is due to a mismatch between tool focus and model card needs: The Model Card Toolkit is useful to report statistics and evaluation metrics directly from the source code (similar to experiment tracking tools like MLflow\footnote{https://mlflow.org/} or Neptune\footnote{https://neptune.ai/home}), but it provides little value for the many other important sections in model cards that require users to manually provide information about intentions, concerns, and ethical deliberations. 

We build on the insight that \textbf{nudging and traceability in an integrated development environment afford better documentation practices.} In our user study, we observed that when our tool was not available, data scientists were overwhelmed with navigating and documenting details of the model development code, leaving no room for considering the model cards, despite the awareness of such a recommendation. In comparison, when the documentation environment was integrated with the coding environment in a meaningful way, data scientists were devoting more effort to improving documentation quality and maintainability. They approached documentation in a more iterative manner and actively used and maintained the trace links between documentation and the source code. The data scientists further spend more time considering the context and impact of the model development and deployment when the explanation and examples of model card sections are nudged in their model development environment. 
We argue that those activities are critical to the comprehensiveness of model documentation, in particular along the ethical axis. 

Among the major features of \toolname{}, nudging is a familiar concept for the CHI community and has been already equipped in the digital interface designers' toolbox. Recent work by Caraban et al. categorizes nudging mechanisms in technology design for health, sustainability, and privacy~\cite{chi2019_23ways_nudge}. The six categories are facilitate, confront, deceive, social influence, fear, and reinforce. Among them, facilitate (e.g., default options) and reinforce (e.g., just-in-time prompts) have been adopted for encouraging software engineering behaviors~\cite{brown2021nudging, kramer1999api}. Similarly, our work uses \textit{facilitate} and \textit{reinforce} mechanisms in the documentation tool for the model developers to more consciously consider and document the model usage and ethical issues during development. We invite researchers to investigate the potential of other mechanisms (such as \textit{confront} by reminding the consequences of not completing the documentation sections) to extend the nudging effect into a broader context.

The concept of traceability and how to approach it through design, however, are less discussed in the human-computer interaction literature. In the context of software development, traceability is an important property for developing safety-critical software~\cite{cleland2012software}. It is often required by regulatory bodies to demonstrate the quality of the software development process and the resulting software. Traceability is also suggested to improve AI accountability by~\citet{10.1145/3351095.3372873}. In our tool \toolname{}, explicit traceability links between code and documentation can support examining the consistency between those two sets of artifacts and therefore improve the accuracy of the documentation and the accountability of the machine learning models. Despite not using the same terminology of traceability, recent work on data documentation also proposes to treat ``text-data connections as persistent, interactive, first class objects''~\cite{chen2022corssdata}. We hope our work and similar attempts can draw more attention from the community to understanding and making use of traceability links during documentation activities that can greatly contribute to the quality and accountability of ML documentation and their systems as a whole.

\subsection{Limitation and Future Work}
As discussed in Section \ref{subsec:user_study_design}, our user study is limited by the lab setting with concrete instructions on the model development tasks that are small in scope. Moreover, we did not explicitly ask the participants about their thoughts on the model cards proposal, but simply informed them about this proposal through the pre-study survey. In future work, we plan to study the impact of \toolname{} in practice with more complex development tasks and other practical concerns (such as model domain and team culture). It is also important to consider how the documentation support should evolve as data scientists gain experience with the model cards proposal. Additionally, while our current work focuses on supporting documentation tasks for data scientists in the notebook environment, future work can be expanded on streamlining documentation activities for their entire workflow that rely on various tools across different environments. 

In our study, we observed how data scientists face difficulty in interpreting the implications of ML models outside the scope of the model development pipeline. This challenge can be amplified in practice. When building ML-enabled software products, an ML model contributes to the overall product but is just one component among many in a system. Such products are typically built by interdisciplinary teams, where software engineers and UI experts integrate ML models into a larger software project, considering many non-ML concerns. Previous work suggested that, in many projects, data scientists building the model are siloed off and rarely interact with teams using their models, causing many conflicts and misunderstandings at the interface between the teams~\cite{NZLK:ICSE22}. Modern MLOps practices and the corresponding rise of the role of ML engineers~\cite{sculley2015hidden,lakshmanan2020machine}, who focus primarily on automating machine learning pipelines, supporting experimentation, and deploying and updating models, introduce further complexities and roles. Model cards, in this case, should not be authored by the data scientists alone. Instead, they should serve as a shared important artifact between the teams that captures the wide range of concerns. In future work, we will examine the potential of considering model cards as  \textit{boundary objects} that are used to negotiate and communicate between data scientists and software engineers to support collaborative interdisciplinary work~\cite{star1989institutional,lee2005between}.

\section{Conclusion}
In this work, we investigated how publicly available ML models are documented, especially when they adopted the model cards proposal. Our assessment of those model cards reveals a clear gap between the proposal and practice. In an effort to move the needle towards meaningful adoption of the model cards proposal and improving documentation practice, we proposed a model documentation tool \toolname{} for data scientists using computational notebooks. As demonstrated in the user study, when the \toolname{} was presented, data scientists more actively improved documentation and considered ethical implications during model development. They also spent considerable effort on the construction and maintenance of trace links between documentation and source code, which supports model accountability. Our work highlights the new opportunities of designing machine learning documentation for long-term benefit through nudging and traceability.

\begin{acks}
We acknowledge the support of the Natural Sciences and
Engineering Research Council of Canada (NSERC) RGPIN-
2021-03538, the National Science Foundation (NSF) award 1813598 and 2131477, and Fonds de recherche
du Québec (FRQNT) 2021-NC-284820. Coursey participated in this work as part of the NSF-supported REU-SE program. We thank our anonymous participants. We also thank our lab members and the CHI reviewers for their constructive feedback.
\end{acks}

\bibliographystyle{ACM-Reference-Format}
\bibliography{reference}


\begin{thebibliography}{71}

%%% the Web address.

\ifx \showCODEN    \undefined \def \showCODEN     #1{\unskip}     \fi
\ifx \showDOI      \undefined \def \showDOI       #1{#1}\fi
\ifx \showISBNx    \undefined \def \showISBNx     #1{\unskip}     \fi
\ifx \showISBNxiii \undefined \def \showISBNxiii  #1{\unskip}     \fi
\ifx \showISSN     \undefined \def \showISSN      #1{\unskip}     \fi
\ifx \showLCCN     \undefined \def \showLCCN      #1{\unskip}     \fi
\ifx \shownote     \undefined \def \shownote      #1{#1}          \fi
\ifx \showarticletitle \undefined \def \showarticletitle #1{#1}   \fi
\ifx \showURL      \undefined \def \showURL       {\relax}        \fi
% invisible to TeX
\providecommand\bibfield[2]{#2}
\providecommand\bibinfo[2]{#2}
\providecommand\natexlab[1]{#1}
\providecommand\showeprint[2][]{arXiv:#2}

\bibitem[\protect\citeauthoryear{Aghajani, Nagy, Linares-V\'{a}squez, Moreno,
  Bavota, Lanza, and Shepherd}{Aghajani et~al\mbox{.}}{2020}]%
        {doc_practitioners_perspective}
\bibfield{author}{\bibinfo{person}{Emad Aghajani}, \bibinfo{person}{Csaba
  Nagy}, \bibinfo{person}{Mario Linares-V\'{a}squez}, \bibinfo{person}{Laura
  Moreno}, \bibinfo{person}{Gabriele Bavota}, \bibinfo{person}{Michele Lanza},
  {and} \bibinfo{person}{David~C. Shepherd}.} \bibinfo{year}{2020}\natexlab{}.
\newblock \showarticletitle{Software Documentation: The Practitioners'
  Perspective}. In \bibinfo{booktitle}{\emph{Proceedings of the ACM/IEEE 42nd
  International Conference on Software Engineering}} (Seoul, South Korea)
  \emph{(\bibinfo{series}{ICSE '20})}. \bibinfo{publisher}{Association for
  Computing Machinery}, \bibinfo{address}{New York, NY, USA},
  \bibinfo{pages}{590–601}.
\newblock
\showISBNx{9781450371216}
\urldef\tempurl%
\url{https://doi.org/10.1145/3377811.3380405}
\showDOI{\tempurl}


\bibitem[\protect\citeauthoryear{Aghajani, Nagy, Vega-M\'{a}rquez,
  Linares-V\'{a}squez, Moreno, Bavota, and Lanza}{Aghajani
  et~al\mbox{.}}{2019}]%
        {aghajani2019software}
\bibfield{author}{\bibinfo{person}{Emad Aghajani}, \bibinfo{person}{Csaba
  Nagy}, \bibinfo{person}{Olga~Lucero Vega-M\'{a}rquez}, \bibinfo{person}{Mario
  Linares-V\'{a}squez}, \bibinfo{person}{Laura Moreno},
  \bibinfo{person}{Gabriele Bavota}, {and} \bibinfo{person}{Michele Lanza}.}
  \bibinfo{year}{2019}\natexlab{}.
\newblock \showarticletitle{Software Documentation Issues Unveiled}. In
  \bibinfo{booktitle}{\emph{Proceedings of the 41st International Conference on
  Software Engineering}} (Montreal, Quebec, Canada)
  \emph{(\bibinfo{series}{ICSE '19})}. \bibinfo{publisher}{IEEE Press},
  \bibinfo{pages}{1199–1210}.
\newblock
\urldef\tempurl%
\url{https://doi.org/10.1109/ICSE.2019.00122}
\showDOI{\tempurl}


\bibitem[\protect\citeauthoryear{Amershi, Begel, Bird, DeLine, Gall, Kamar,
  Nagappan, Nushi, and Zimmermann}{Amershi et~al\mbox{.}}{2019}]%
        {se4ml_case_study}
\bibfield{author}{\bibinfo{person}{Saleema Amershi}, \bibinfo{person}{Andrew
  Begel}, \bibinfo{person}{Christian Bird}, \bibinfo{person}{Robert DeLine},
  \bibinfo{person}{Harald Gall}, \bibinfo{person}{Ece Kamar},
  \bibinfo{person}{Nachiappan Nagappan}, \bibinfo{person}{Besmira Nushi}, {and}
  \bibinfo{person}{Thomas Zimmermann}.} \bibinfo{year}{2019}\natexlab{}.
\newblock \showarticletitle{Software Engineering for Machine Learning: A Case
  Study}. In \bibinfo{booktitle}{\emph{Proceedings of the 41st International
  Conference on Software Engineering: Software Engineering in Practice}}
  (Montreal, Quebec, Canada) \emph{(\bibinfo{series}{ICSE-SEIP '19})}.
  \bibinfo{publisher}{IEEE Press}, \bibinfo{pages}{291–300}.
\newblock
\urldef\tempurl%
\url{https://doi.org/10.1109/ICSE-SEIP.2019.00042}
\showDOI{\tempurl}


\bibitem[\protect\citeauthoryear{Arnold, Bellamy, Hind, Houde, Mehta,
  Mojsilovi{\'c}, Nair, Ramamurthy, Olteanu, Piorkowski, et~al\mbox{.}}{Arnold
  et~al\mbox{.}}{2019}]%
        {arnold2019factsheets}
\bibfield{author}{\bibinfo{person}{Matthew Arnold}, \bibinfo{person}{Rachel~KE
  Bellamy}, \bibinfo{person}{Michael Hind}, \bibinfo{person}{Stephanie Houde},
  \bibinfo{person}{Sameep Mehta}, \bibinfo{person}{Aleksandra Mojsilovi{\'c}},
  \bibinfo{person}{Ravi Nair}, \bibinfo{person}{K~Natesan Ramamurthy},
  \bibinfo{person}{Alexandra Olteanu}, \bibinfo{person}{David Piorkowski},
  {et~al\mbox{.}}} \bibinfo{year}{2019}\natexlab{}.
\newblock \showarticletitle{FactSheets: Increasing trust in AI services through
  supplier's declarations of conformity}.
\newblock \bibinfo{journal}{\emph{IBM Journal of Research and Development}}
  \bibinfo{volume}{63}, \bibinfo{number}{4/5} (\bibinfo{year}{2019}),
  \bibinfo{pages}{6--1}.
\newblock


\bibitem[\protect\citeauthoryear{Arya, Guo, and Robillard}{Arya
  et~al\mbox{.}}{2020}]%
        {arya2020information}
\bibfield{author}{\bibinfo{person}{Deeksha~M Arya}, \bibinfo{person}{Jin~LC
  Guo}, {and} \bibinfo{person}{Martin~P Robillard}.}
  \bibinfo{year}{2020}\natexlab{}.
\newblock \showarticletitle{Information correspondence between types of
  documentation for APIs}.
\newblock \bibinfo{journal}{\emph{Empirical Software Engineering}}
  \bibinfo{volume}{25}, \bibinfo{number}{5} (\bibinfo{year}{2020}),
  \bibinfo{pages}{4069--4096}.
\newblock


\bibitem[\protect\citeauthoryear{Blandford, Cox, and Cairns}{Blandford
  et~al\mbox{.}}{2008}]%
        {blandford2008controlled}
\bibfield{author}{\bibinfo{person}{Ann Blandford}, \bibinfo{person}{Anna~L.
  Cox}, {and} \bibinfo{person}{Paul Cairns}.} \bibinfo{year}{2008}\natexlab{}.
\newblock \bibinfo{booktitle}{\emph{Controlled experiments}}.
\newblock \bibinfo{publisher}{Cambridge University Press},
  \bibinfo{pages}{1–16}.
\newblock
\urldef\tempurl%
\url{https://doi.org/10.1017/CBO9780511814570.002}
\showDOI{\tempurl}


\bibitem[\protect\citeauthoryear{Blodgett, Barocas, Daum{\'e}~III, and
  Wallach}{Blodgett et~al\mbox{.}}{2020}]%
        {blodgett-etal-2020-language}
\bibfield{author}{\bibinfo{person}{Su~Lin Blodgett}, \bibinfo{person}{Solon
  Barocas}, \bibinfo{person}{Hal Daum{\'e}~III}, {and} \bibinfo{person}{Hanna
  Wallach}.} \bibinfo{year}{2020}\natexlab{}.
\newblock \showarticletitle{Language (Technology) is Power: A Critical Survey
  of {``}Bias{''} in {NLP}}. In \bibinfo{booktitle}{\emph{Proceedings of the
  58th Annual Meeting of the Association for Computational Linguistics}}.
  \bibinfo{publisher}{Association for Computational Linguistics},
  \bibinfo{address}{Online}, \bibinfo{pages}{5454--5476}.
\newblock
\urldef\tempurl%
\url{https://doi.org/10.18653/v1/2020.acl-main.485}
\showDOI{\tempurl}


\bibitem[\protect\citeauthoryear{Boyd}{Boyd}{2021}]%
        {boyd2021datasheets}
\bibfield{author}{\bibinfo{person}{Karen~L Boyd}.}
  \bibinfo{year}{2021}\natexlab{}.
\newblock \showarticletitle{Datasheets for Datasets help ML Engineers Notice
  and Understand Ethical Issues in Training Data}.
\newblock \bibinfo{journal}{\emph{Proceedings of the ACM on Human-Computer
  Interaction}} \bibinfo{volume}{5}, \bibinfo{number}{CSCW2}
  (\bibinfo{year}{2021}), \bibinfo{pages}{1--27}.
\newblock


\bibitem[\protect\citeauthoryear{Brown and Parnin}{Brown and Parnin}{2021}]%
        {brown2021nudging}
\bibfield{author}{\bibinfo{person}{Chris Brown} {and} \bibinfo{person}{Chris
  Parnin}.} \bibinfo{year}{2021}\natexlab{}.
\newblock \showarticletitle{Nudging Students Toward Better Software Engineering
  Behaviors}. In \bibinfo{booktitle}{\emph{2021 IEEE/ACM Third International
  Workshop on Bots in Software Engineering (BotSE)}}. \bibinfo{publisher}{IEEE
  Press}, \bibinfo{pages}{11--15}.
\newblock
\urldef\tempurl%
\url{https://doi.org/10.1109/BotSE52550.2021.00010}
\showDOI{\tempurl}


\bibitem[\protect\citeauthoryear{Buolamwini and Gebru}{Buolamwini and
  Gebru}{2018}]%
        {buolamwini2018gender}
\bibfield{author}{\bibinfo{person}{Joy Buolamwini} {and}
  \bibinfo{person}{Timnit Gebru}.} \bibinfo{year}{2018}\natexlab{}.
\newblock \showarticletitle{Gender Shades: Intersectional Accuracy Disparities
  in Commercial Gender Classification}. In
  \bibinfo{booktitle}{\emph{Proceedings of the 1st Conference on Fairness,
  Accountability and Transparency}} \emph{(\bibinfo{series}{Proceedings of
  Machine Learning Research}, Vol.~\bibinfo{volume}{81})},
  \bibfield{editor}{\bibinfo{person}{Sorelle~A. Friedler} {and}
  \bibinfo{person}{Christo Wilson}} (Eds.). \bibinfo{publisher}{PMLR},
  \bibinfo{pages}{77--91}.
\newblock
\urldef\tempurl%
\url{https://proceedings.mlr.press/v81/buolamwini18a.html}
\showURL{%
\tempurl}


\bibitem[\protect\citeauthoryear{Caraban, Karapanos, Gon\c{c}alves, and
  Campos}{Caraban et~al\mbox{.}}{2019}]%
        {chi2019_23ways_nudge}
\bibfield{author}{\bibinfo{person}{Ana Caraban}, \bibinfo{person}{Evangelos
  Karapanos}, \bibinfo{person}{Daniel Gon\c{c}alves}, {and}
  \bibinfo{person}{Pedro Campos}.} \bibinfo{year}{2019}\natexlab{}.
\newblock \bibinfo{booktitle}{\emph{23 Ways to Nudge: A Review of
  Technology-Mediated Nudging in Human-Computer Interaction}}.
\newblock \bibinfo{publisher}{Association for Computing Machinery},
  \bibinfo{address}{New York, NY, USA}, \bibinfo{pages}{1–15}.
\newblock
\showISBNx{9781450359702}
\urldef\tempurl%
\url{https://doi.org/10.1145/3290605.3300733}
\showURL{%
\tempurl}


\bibitem[\protect\citeauthoryear{Chattopadhyay, Prasad, Henley, Sarma, and
  Barik}{Chattopadhyay et~al\mbox{.}}{2020}]%
        {souti2020notebook}
\bibfield{author}{\bibinfo{person}{Souti Chattopadhyay},
  \bibinfo{person}{Ishita Prasad}, \bibinfo{person}{Austin~Z. Henley},
  \bibinfo{person}{Anita Sarma}, {and} \bibinfo{person}{Titus Barik}.}
  \bibinfo{year}{2020}\natexlab{}.
\newblock \showarticletitle{What's Wrong with Computational Notebooks? Pain
  Points, Needs, and Design Opportunities}. In
  \bibinfo{booktitle}{\emph{Proceedings of the 2020 CHI Conference on Human
  Factors in Computing Systems}} (Honolulu, HI, USA)
  \emph{(\bibinfo{series}{CHI '20})}. \bibinfo{publisher}{Association for
  Computing Machinery}, \bibinfo{address}{New York, NY, USA},
  \bibinfo{pages}{1–12}.
\newblock
\showISBNx{9781450367080}
\urldef\tempurl%
\url{https://doi.org/10.1145/3313831.3376729}
\showDOI{\tempurl}


\bibitem[\protect\citeauthoryear{Chen and Xia}{Chen and Xia}{2022}]%
        {chen2022corssdata}
\bibfield{author}{\bibinfo{person}{Zhutian Chen} {and} \bibinfo{person}{Haijun
  Xia}.} \bibinfo{year}{2022}\natexlab{}.
\newblock \showarticletitle{CrossData: Leveraging Text-Data Connections for
  Authoring Data Documents}. In \bibinfo{booktitle}{\emph{Proceedings of the
  2022 CHI Conference on Human Factors in Computing Systems}} (New Orleans, LA,
  USA) \emph{(\bibinfo{series}{CHI '22})}. \bibinfo{publisher}{Association for
  Computing Machinery}, \bibinfo{address}{New York, NY, USA}, Article
  \bibinfo{articleno}{95}, \bibinfo{numpages}{15}~pages.
\newblock
\showISBNx{9781450391573}
\urldef\tempurl%
\url{https://doi.org/10.1145/3491102.3517485}
\showDOI{\tempurl}


\bibitem[\protect\citeauthoryear{Cohen}{Cohen}{1960}]%
        {cohen1960kappa}
\bibfield{author}{\bibinfo{person}{Jacob Cohen}.}
  \bibinfo{year}{1960}\natexlab{}.
\newblock \showarticletitle{A Coefficient of Agreement for Nominal Scales}.
\newblock \bibinfo{journal}{\emph{Educational and Psychological Measurement}}
  \bibinfo{volume}{20}, \bibinfo{number}{1} (\bibinfo{year}{1960}),
  \bibinfo{pages}{37--46}.
\newblock
\urldef\tempurl%
\url{https://doi.org/10.1177/001316446002000104}
\showDOI{\tempurl}
\showeprint{https://doi.org/10.1177/001316446002000104}


\bibitem[\protect\citeauthoryear{Cortés-Coy, Linares-Vásquez, Aponte, and
  Poshyvanyk}{Cortés-Coy et~al\mbox{.}}{2014}]%
        {6975661}
\bibfield{author}{\bibinfo{person}{Luis~Fernando Cortés-Coy},
  \bibinfo{person}{Mario Linares-Vásquez}, \bibinfo{person}{Jairo Aponte},
  {and} \bibinfo{person}{Denys Poshyvanyk}.} \bibinfo{year}{2014}\natexlab{}.
\newblock \showarticletitle{On Automatically Generating Commit Messages via
  Summarization of Source Code Changes}. In \bibinfo{booktitle}{\emph{2014 IEEE
  14th International Working Conference on Source Code Analysis and
  Manipulation}}. \bibinfo{publisher}{IEEE Press}, \bibinfo{pages}{275--284}.
\newblock
\urldef\tempurl%
\url{https://doi.org/10.1109/SCAM.2014.14}
\showDOI{\tempurl}


\bibitem[\protect\citeauthoryear{Crisan, Drouhard, Vig, and Rajani}{Crisan
  et~al\mbox{.}}{2022}]%
        {crisan_interactive_2022}
\bibfield{author}{\bibinfo{person}{Anamaria Crisan}, \bibinfo{person}{Margaret
  Drouhard}, \bibinfo{person}{Jesse Vig}, {and} \bibinfo{person}{Nazneen
  Rajani}.} \bibinfo{year}{2022}\natexlab{}.
\newblock \showarticletitle{Interactive {Model} {Cards}: {A} {Human}-{Centered}
  {Approach} to {Model} {Documentation}}. In \bibinfo{booktitle}{\emph{2022
  {ACM} {Conference} on {Fairness}, {Accountability}, and {Transparency}}}.
  \bibinfo{publisher}{ACM}, \bibinfo{address}{Seoul Republic of Korea},
  \bibinfo{pages}{427--439}.
\newblock
\showISBNx{978-1-4503-9352-2}
\urldef\tempurl%
\url{https://doi.org/10.1145/3531146.3533108}
\showDOI{\tempurl}


\bibitem[\protect\citeauthoryear{Dastin}{Dastin}{2018}]%
        {amazon_recruiting_tool}
\bibfield{author}{\bibinfo{person}{Jeffrey Dastin}.}
  \bibinfo{year}{2018}\natexlab{}.
\newblock \bibinfo{booktitle}{\emph{Amazon scraps secret AI recruiting tool
  that showed bias against women}}.
\newblock
\urldef\tempurl%
\url{https://www.reuters.com/article/us-amazon-com-jobs-automation-insight-idUSKCN1MK08G}
\showURL{%
\tempurl}


\bibitem[\protect\citeauthoryear{developers}{developers}{[n.d.]}]%
        {sphinx}
\bibfield{author}{\bibinfo{person}{The~Sphinx developers}.}
  \bibinfo{year}{[n.d.]}\natexlab{}.
\newblock \bibinfo{booktitle}{\emph{Welcome: Sphinx makes it easy to create
  intelligent and beautiful documentation.}}
\newblock


\bibitem[\protect\citeauthoryear{Dressel and Farid}{Dressel and Farid}{2018}]%
        {dressel2018accuracy}
\bibfield{author}{\bibinfo{person}{Julia Dressel} {and} \bibinfo{person}{Hany
  Farid}.} \bibinfo{year}{2018}\natexlab{}.
\newblock \showarticletitle{The accuracy, fairness, and limits of predicting
  recidivism}.
\newblock \bibinfo{journal}{\emph{Science advances}} \bibinfo{volume}{4},
  \bibinfo{number}{1} (\bibinfo{year}{2018}), \bibinfo{pages}{eaao5580}.
\newblock


\bibitem[\protect\citeauthoryear{Drosos, Barik, Guo, DeLine, and
  Gulwani}{Drosos et~al\mbox{.}}{2020}]%
        {ian2022wrex}
\bibfield{author}{\bibinfo{person}{Ian Drosos}, \bibinfo{person}{Titus Barik},
  \bibinfo{person}{Philip~J. Guo}, \bibinfo{person}{Robert DeLine}, {and}
  \bibinfo{person}{Sumit Gulwani}.} \bibinfo{year}{2020}\natexlab{}.
\newblock \bibinfo{booktitle}{\emph{Wrex: A Unified Programming-by-Example
  Interaction for Synthesizing Readable Code for Data Scientists}}.
\newblock \bibinfo{publisher}{Association for Computing Machinery},
  \bibinfo{address}{New York, NY, USA}, \bibinfo{pages}{1–12}.
\newblock
\showISBNx{9781450367080}
\urldef\tempurl%
\url{https://doi.org/10.1145/3313831.3376442}
\showURL{%
\tempurl}


\bibitem[\protect\citeauthoryear{Face}{Face}{2021a}]%
        {huggingface-modelcarddocs}
\bibfield{author}{\bibinfo{person}{Hugging Face}.}
  \bibinfo{year}{2021}\natexlab{a}.
\newblock \bibinfo{title}{Hugging Face Documentation: Building a model card}.
\newblock
\newblock
\urldef\tempurl%
\url{https://huggingface.co/course/chapter4/4}
\showURL{%
Retrieved Jan 5, 2022 from \tempurl}


\bibitem[\protect\citeauthoryear{Face}{Face}{2021b}]%
        {huggingface-modeldocs}
\bibfield{author}{\bibinfo{person}{Hugging Face}.}
  \bibinfo{year}{2021}\natexlab{b}.
\newblock \bibinfo{title}{Hugging Face Documentation: Model Repos docs}.
\newblock
\newblock
\urldef\tempurl%
\url{https://huggingface.co/docs/hub/model-repos}
\showURL{%
Retrieved Jan 5, 2022 from \tempurl}


\bibitem[\protect\citeauthoryear{Face}{Face}{2021c}]%
        {huggingface}
\bibfield{author}{\bibinfo{person}{Hugging Face}.}
  \bibinfo{year}{2021}\natexlab{c}.
\newblock \bibinfo{title}{Hugging Face – The AI community building the
  future.}
\newblock
\newblock
\urldef\tempurl%
\url{https://huggingface.co/}
\showURL{%
Retrieved July 20, 2021 from \tempurl}


\bibitem[\protect\citeauthoryear{Gebru, Morgenstern, Vecchione, Vaughan,
  Wallach, III, and Crawford}{Gebru et~al\mbox{.}}{2021}]%
        {gebru2018datasheets}
\bibfield{author}{\bibinfo{person}{Timnit Gebru}, \bibinfo{person}{Jamie
  Morgenstern}, \bibinfo{person}{Briana Vecchione},
  \bibinfo{person}{Jennifer~Wortman Vaughan}, \bibinfo{person}{Hanna Wallach},
  \bibinfo{person}{Hal~Daum\'{e} III}, {and} \bibinfo{person}{Kate Crawford}.}
  \bibinfo{year}{2021}\natexlab{}.
\newblock \showarticletitle{Datasheets for Datasets}.
\newblock \bibinfo{journal}{\emph{Commun. ACM}} \bibinfo{volume}{64},
  \bibinfo{number}{12} (\bibinfo{date}{nov} \bibinfo{year}{2021}),
  \bibinfo{pages}{86–92}.
\newblock
\showISSN{0001-0782}
\urldef\tempurl%
\url{https://doi.org/10.1145/3458723}
\showDOI{\tempurl}


\bibitem[\protect\citeauthoryear{Giner-Miguelez, G\'{o}mez, and
  Cabot}{Giner-Miguelez et~al\mbox{.}}{2022}]%
        {giner-miguelez_describeml_2017}
\bibfield{author}{\bibinfo{person}{Joan Giner-Miguelez}, \bibinfo{person}{Abel
  G\'{o}mez}, {and} \bibinfo{person}{Jordi Cabot}.}
  \bibinfo{year}{2022}\natexlab{}.
\newblock \showarticletitle{DescribeML: A Tool for Describing Machine Learning
  Datasets}. In \bibinfo{booktitle}{\emph{Proceedings of the 25th International
  Conference on Model Driven Engineering Languages and Systems: Companion
  Proceedings}} (Montreal, Quebec, Canada) \emph{(\bibinfo{series}{MODELS
  '22})}. \bibinfo{publisher}{Association for Computing Machinery},
  \bibinfo{address}{New York, NY, USA}, \bibinfo{pages}{22–26}.
\newblock
\showISBNx{9781450394673}
\urldef\tempurl%
\url{https://doi.org/10.1145/3550356.3559087}
\showDOI{\tempurl}


\bibitem[\protect\citeauthoryear{Google}{Google}{2020}]%
        {modelcardtoolkit}
\bibfield{author}{\bibinfo{person}{Google}.} \bibinfo{year}{2020}\natexlab{}.
\newblock \bibinfo{title}{model-card-toolkit}.
\newblock
  \bibinfo{howpublished}{\url{https://github.com/tensorflow/model-card-toolkit}}.
\newblock


\bibitem[\protect\citeauthoryear{Gotel, Cleland-Huang, Hayes, Zisman, Egyed,
  Gr{\"u}nbacher, Dekhtyar, Antoniol, Maletic, and M{\"a}der}{Gotel
  et~al\mbox{.}}{2012}]%
        {cleland2012software}
\bibfield{author}{\bibinfo{person}{Orlena Gotel}, \bibinfo{person}{Jane
  Cleland-Huang}, \bibinfo{person}{Jane~Huffman Hayes}, \bibinfo{person}{Andrea
  Zisman}, \bibinfo{person}{Alexander Egyed}, \bibinfo{person}{Paul
  Gr{\"u}nbacher}, \bibinfo{person}{Alex Dekhtyar}, \bibinfo{person}{Giuliano
  Antoniol}, \bibinfo{person}{Jonathan Maletic}, {and} \bibinfo{person}{Patrick
  M{\"a}der}.} \bibinfo{year}{2012}\natexlab{}.
\newblock \bibinfo{booktitle}{\emph{Traceability Fundamentals}}.
\newblock \bibinfo{publisher}{Springer London}, \bibinfo{address}{London},
  \bibinfo{pages}{3--22}.
\newblock
\showISBNx{978-1-4471-2239-5}
\urldef\tempurl%
\url{https://doi.org/10.1007/978-1-4471-2239-5_1}
\showDOI{\tempurl}


\bibitem[\protect\citeauthoryear{Haakman, Cruz, Huijgens, and van
  Deursen}{Haakman et~al\mbox{.}}{2021}]%
        {10.1007/s10664-021-09993-1}
\bibfield{author}{\bibinfo{person}{Mark Haakman}, \bibinfo{person}{Lu\'{\i}s
  Cruz}, \bibinfo{person}{Hennie Huijgens}, {and} \bibinfo{person}{Arie van
  Deursen}.} \bibinfo{year}{2021}\natexlab{}.
\newblock \showarticletitle{AI Lifecycle Models Need to Be Revised: An
  Exploratory Study in Fintech}.
\newblock \bibinfo{journal}{\emph{Empirical Softw. Engg.}}
  \bibinfo{volume}{26}, \bibinfo{number}{5} (\bibinfo{date}{sep}
  \bibinfo{year}{2021}), 29.
\newblock
\showISSN{1382-3256}
\urldef\tempurl%
\url{https://doi.org/10.1007/s10664-021-09993-1}
\showDOI{\tempurl}


\bibitem[\protect\citeauthoryear{Head, Hohman, Barik, Drucker, and DeLine}{Head
  et~al\mbox{.}}{2019a}]%
        {head2019managing}
\bibfield{author}{\bibinfo{person}{Andrew Head}, \bibinfo{person}{Fred Hohman},
  \bibinfo{person}{Titus Barik}, \bibinfo{person}{Steven~M. Drucker}, {and}
  \bibinfo{person}{Robert DeLine}.} \bibinfo{year}{2019}\natexlab{a}.
\newblock \showarticletitle{Managing Messes in Computational Notebooks}. In
  \bibinfo{booktitle}{\emph{Proceedings of the 2019 CHI Conference on Human
  Factors in Computing Systems}} (Glasgow, Scotland Uk)
  \emph{(\bibinfo{series}{CHI '19})}. \bibinfo{publisher}{Association for
  Computing Machinery}, \bibinfo{address}{New York, NY, USA},
  \bibinfo{pages}{1–12}.
\newblock
\showISBNx{9781450359702}
\urldef\tempurl%
\url{https://doi.org/10.1145/3290605.3300500}
\showDOI{\tempurl}


\bibitem[\protect\citeauthoryear{Head, Hohman, Barik, Drucker, and DeLine}{Head
  et~al\mbox{.}}{2019b}]%
        {gather2019andrewhead}
\bibfield{author}{\bibinfo{person}{Andrew Head}, \bibinfo{person}{Fred Hohman},
  \bibinfo{person}{Titus Barik}, \bibinfo{person}{Steven~M. Drucker}, {and}
  \bibinfo{person}{Robert DeLine}.} \bibinfo{year}{2019}\natexlab{b}.
\newblock \showarticletitle{Managing Messes in Computational Notebooks}. In
  \bibinfo{booktitle}{\emph{Proceedings of the 2019 CHI Conference on Human
  Factors in Computing Systems}} (Glasgow, Scotland Uk)
  \emph{(\bibinfo{series}{CHI '19})}. \bibinfo{publisher}{Association for
  Computing Machinery}, \bibinfo{address}{New York, NY, USA},
  \bibinfo{pages}{1–12}.
\newblock
\showISBNx{9781450359702}
\urldef\tempurl%
\url{https://doi.org/10.1145/3290605.3300500}
\showDOI{\tempurl}


\bibitem[\protect\citeauthoryear{Head, Jiang, Smith, Hearst, and Hartmann}{Head
  et~al\mbox{.}}{2020}]%
        {torii}
\bibfield{author}{\bibinfo{person}{Andrew Head}, \bibinfo{person}{Jason Jiang},
  \bibinfo{person}{James Smith}, \bibinfo{person}{Marti~A. Hearst}, {and}
  \bibinfo{person}{Bj\"{o}rn Hartmann}.} \bibinfo{year}{2020}\natexlab{}.
\newblock \showarticletitle{Composing Flexibly-Organized Step-by-Step Tutorials
  from Linked Source Code, Snippets, and Outputs}. In
  \bibinfo{booktitle}{\emph{Proceedings of the 2020 CHI Conference on Human
  Factors in Computing Systems}} (Honolulu, HI, USA)
  \emph{(\bibinfo{series}{CHI '20})}. \bibinfo{publisher}{Association for
  Computing Machinery}, \bibinfo{address}{New York, NY, USA},
  \bibinfo{pages}{1–12}.
\newblock
\showISBNx{9781450367080}
\urldef\tempurl%
\url{https://doi.org/10.1145/3313831.3376798}
\showDOI{\tempurl}


\bibitem[\protect\citeauthoryear{Hellman, Jang, Treude, Huang, and Guo}{Hellman
  et~al\mbox{.}}{2021}]%
        {hellman2021generating}
\bibfield{author}{\bibinfo{person}{Jazlyn Hellman}, \bibinfo{person}{Eunbee
  Jang}, \bibinfo{person}{Christoph Treude}, \bibinfo{person}{Chenzhun Huang},
  {and} \bibinfo{person}{Jin~LC Guo}.} \bibinfo{year}{2021}\natexlab{}.
\newblock \showarticletitle{Generating GitHub Repository Descriptions: A
  Comparison of Manual and Automated Approaches}.
\newblock  (\bibinfo{year}{2021}).
\newblock
\urldef\tempurl%
\url{https://doi.org/10.48550/arXiv.2110.13283}
\showDOI{\tempurl}
\showeprint{arXiv:2110.13283}


\bibitem[\protect\citeauthoryear{Holland, Hosny, and Newman}{Holland
  et~al\mbox{.}}{2020}]%
        {holland2020dataset}
\bibfield{author}{\bibinfo{person}{Sarah Holland}, \bibinfo{person}{Ahmed
  Hosny}, {and} \bibinfo{person}{Sarah Newman}.}
  \bibinfo{year}{2020}\natexlab{}.
\newblock \showarticletitle{The dataset nutrition label}.
\newblock \bibinfo{journal}{\emph{Data Protection and Privacy, Volume 12: Data
  Protection and Democracy}}  \bibinfo{volume}{12} (\bibinfo{year}{2020}),
  \bibinfo{pages}{1}.
\newblock


\bibitem[\protect\citeauthoryear{Holstein, Wortman~Vaughan, Daum\'{e}, Dudik,
  and Wallach}{Holstein et~al\mbox{.}}{2019}]%
        {10.1145/3290605.3300830}
\bibfield{author}{\bibinfo{person}{Kenneth Holstein}, \bibinfo{person}{Jennifer
  Wortman~Vaughan}, \bibinfo{person}{Hal Daum\'{e}}, \bibinfo{person}{Miro
  Dudik}, {and} \bibinfo{person}{Hanna Wallach}.}
  \bibinfo{year}{2019}\natexlab{}.
\newblock \bibinfo{booktitle}{\emph{Improving Fairness in Machine Learning
  Systems: What Do Industry Practitioners Need?}}
\newblock \bibinfo{publisher}{Association for Computing Machinery},
  \bibinfo{address}{New York, NY, USA}, \bibinfo{pages}{1–16}.
\newblock
\showISBNx{9781450359702}
\urldef\tempurl%
\url{https://doi.org/10.1145/3290605.3300830}
\showURL{%
\tempurl}


\bibitem[\protect\citeauthoryear{Hopkins and Booth}{Hopkins and Booth}{2021}]%
        {hopkins2021machine}
\bibfield{author}{\bibinfo{person}{Aspen Hopkins} {and} \bibinfo{person}{Serena
  Booth}.} \bibinfo{year}{2021}\natexlab{}.
\newblock \showarticletitle{Machine Learning Practices Outside Big Tech: How
  Resource Constraints Challenge Responsible Development}. In
  \bibinfo{booktitle}{\emph{Proceedings of the 2021 AAAI/ACM Conference on AI,
  Ethics, and Society}} (Virtual Event, USA) \emph{(\bibinfo{series}{AIES
  '21})}. \bibinfo{publisher}{Association for Computing Machinery},
  \bibinfo{address}{New York, NY, USA}, \bibinfo{pages}{134–145}.
\newblock
\showISBNx{9781450384735}
\urldef\tempurl%
\url{https://doi.org/10.1145/3461702.3462527}
\showDOI{\tempurl}


\bibitem[\protect\citeauthoryear{Hulten}{Hulten}{2018}]%
        {hulten2018building}
\bibfield{author}{\bibinfo{person}{Geoff Hulten}.}
  \bibinfo{year}{2018}\natexlab{}.
\newblock \bibinfo{booktitle}{\emph{Building Intelligent Systems: A Guide to
  Machine Learning Engineering}}.
\newblock \bibinfo{publisher}{Apress}.
\newblock


\bibitem[\protect\citeauthoryear{Jigsaw}{Jigsaw}{2017}]%
        {perspectiveAPI}
\bibfield{author}{\bibinfo{person}{Jigsaw}.} \bibinfo{year}{2017}\natexlab{}.
\newblock \bibinfo{title}{Perspective API}.
\newblock \bibinfo{howpublished}{\url{https://www.perspectiveapi.com/}}.
\newblock


\bibitem[\protect\citeauthoryear{Kery, Radensky, Arya, John, and Myers}{Kery
  et~al\mbox{.}}{2018}]%
        {kery2018story}
\bibfield{author}{\bibinfo{person}{Mary~Beth Kery}, \bibinfo{person}{Marissa
  Radensky}, \bibinfo{person}{Mahima Arya}, \bibinfo{person}{Bonnie~E. John},
  {and} \bibinfo{person}{Brad~A. Myers}.} \bibinfo{year}{2018}\natexlab{}.
\newblock \showarticletitle{The Story in the Notebook: Exploratory Data Science
  Using a Literate Programming Tool}. In \bibinfo{booktitle}{\emph{Proceedings
  of the 2018 CHI Conference on Human Factors in Computing Systems}} (Montreal
  QC, Canada) \emph{(\bibinfo{series}{CHI '18})}.
  \bibinfo{publisher}{Association for Computing Machinery},
  \bibinfo{address}{New York, NY, USA}, \bibinfo{pages}{1–11}.
\newblock
\showISBNx{9781450356206}
\urldef\tempurl%
\url{https://doi.org/10.1145/3173574.3173748}
\showDOI{\tempurl}


\bibitem[\protect\citeauthoryear{Kramer}{Kramer}{1999}]%
        {kramer1999api}
\bibfield{author}{\bibinfo{person}{Douglas Kramer}.}
  \bibinfo{year}{1999}\natexlab{}.
\newblock \showarticletitle{API Documentation from Source Code Comments: A Case
  Study of Javadoc}. In \bibinfo{booktitle}{\emph{Proceedings of the 17th
  Annual International Conference on Computer Documentation}} (New Orleans,
  Louisiana, USA) \emph{(\bibinfo{series}{SIGDOC '99})}.
  \bibinfo{publisher}{Association for Computing Machinery},
  \bibinfo{address}{New York, NY, USA}, \bibinfo{pages}{147–153}.
\newblock
\showISBNx{1581130724}
\urldef\tempurl%
\url{https://doi.org/10.1145/318372.318577}
\showDOI{\tempurl}


\bibitem[\protect\citeauthoryear{Lakshmanan, Robinson, and Munn}{Lakshmanan
  et~al\mbox{.}}{2020}]%
        {lakshmanan2020machine}
\bibfield{author}{\bibinfo{person}{Valliappa Lakshmanan}, \bibinfo{person}{Sara
  Robinson}, {and} \bibinfo{person}{Michael Munn}.}
  \bibinfo{year}{2020}\natexlab{}.
\newblock \bibinfo{booktitle}{\emph{Machine learning design patterns}}.
\newblock \bibinfo{publisher}{O'Reilly Media}.
\newblock


\bibitem[\protect\citeauthoryear{Lee}{Lee}{2005}]%
        {lee2005between}
\bibfield{author}{\bibinfo{person}{Charlotte~P. Lee}.}
  \bibinfo{year}{2005}\natexlab{}.
\newblock \showarticletitle{Between Chaos and Routine: Boundary Negotiating
  Artifacts in Collaboration}. In \bibinfo{booktitle}{\emph{ECSCW 2005}},
  \bibfield{editor}{\bibinfo{person}{Hans Gellersen}, \bibinfo{person}{Kjeld
  Schmidt}, \bibinfo{person}{Michel Beaudouin-Lafon}, {and}
  \bibinfo{person}{Wendy Mackay}} (Eds.). \bibinfo{publisher}{Springer
  Netherlands}, \bibinfo{address}{Dordrecht}, \bibinfo{pages}{387--406}.
\newblock
\showISBNx{978-1-4020-4023-8}


\bibitem[\protect\citeauthoryear{Liu, Luo, Wang, and Tang}{Liu
  et~al\mbox{.}}{2015}]%
        {liu2015deep}
\bibfield{author}{\bibinfo{person}{Z. Liu}, \bibinfo{person}{P. Luo},
  \bibinfo{person}{X. Wang}, {and} \bibinfo{person}{X. Tang}.}
  \bibinfo{year}{2015}\natexlab{}.
\newblock \showarticletitle{Deep Learning Face Attributes in the Wild}. In
  \bibinfo{booktitle}{\emph{2015 IEEE International Conference on Computer
  Vision (ICCV)}}. \bibinfo{publisher}{IEEE Computer Society},
  \bibinfo{address}{Los Alamitos, CA, USA}, \bibinfo{pages}{3730--3738}.
\newblock
\showISSN{2380-7504}
\urldef\tempurl%
\url{https://doi.org/10.1109/ICCV.2015.425}
\showDOI{\tempurl}


\bibitem[\protect\citeauthoryear{Maffey, Dotterrer, Niemann, Cruickshank,
  Lewis, and Kästner}{Maffey et~al\mbox{.}}{2023}]%
        {mlte}
\bibfield{author}{\bibinfo{person}{Katherine~R. Maffey}, \bibinfo{person}{Kyle
  Dotterrer}, \bibinfo{person}{Jennifer Niemann}, \bibinfo{person}{Iain
  Cruickshank}, \bibinfo{person}{Grace~A. Lewis}, {and}
  \bibinfo{person}{Christian Kästner}.} \bibinfo{year}{2023}\natexlab{}.
\newblock \showarticletitle{MLTEing Models: Negotiating, Evaluating, and
  Documenting Model and System Qualities}. In \bibinfo{booktitle}{\emph{Proc.
  International Conference on Software Engineering: New Ideas and Emerging
  Results (ICSE-NIER)}}.
\newblock


\bibitem[\protect\citeauthoryear{McBurney and McMillan}{McBurney and
  McMillan}{2014}]%
        {mcburney2014automatic}
\bibfield{author}{\bibinfo{person}{Paul~W. McBurney} {and}
  \bibinfo{person}{Collin McMillan}.} \bibinfo{year}{2014}\natexlab{}.
\newblock \showarticletitle{Automatic Documentation Generation via Source Code
  Summarization of Method Context}. In \bibinfo{booktitle}{\emph{Proceedings of
  the 22nd International Conference on Program Comprehension}} (Hyderabad,
  India) \emph{(\bibinfo{series}{ICPC 2014})}. \bibinfo{publisher}{Association
  for Computing Machinery}, \bibinfo{address}{New York, NY, USA},
  \bibinfo{pages}{279–290}.
\newblock
\showISBNx{9781450328791}
\urldef\tempurl%
\url{https://doi.org/10.1145/2597008.2597149}
\showDOI{\tempurl}


\bibitem[\protect\citeauthoryear{Mitchell, Wu, Zaldivar, Barnes, Vasserman,
  Hutchinson, Spitzer, Raji, and Gebru}{Mitchell et~al\mbox{.}}{2019}]%
        {modelcard}
\bibfield{author}{\bibinfo{person}{Margaret Mitchell}, \bibinfo{person}{Simone
  Wu}, \bibinfo{person}{Andrew Zaldivar}, \bibinfo{person}{Parker Barnes},
  \bibinfo{person}{Lucy Vasserman}, \bibinfo{person}{Ben Hutchinson},
  \bibinfo{person}{Elena Spitzer}, \bibinfo{person}{Inioluwa~Deborah Raji},
  {and} \bibinfo{person}{Timnit Gebru}.} \bibinfo{year}{2019}\natexlab{}.
\newblock \showarticletitle{Model Cards for Model Reporting}. In
  \bibinfo{booktitle}{\emph{Proceedings of the Conference on Fairness,
  Accountability, and Transparency}} (Atlanta, GA, USA)
  \emph{(\bibinfo{series}{FAT* '19})}. \bibinfo{publisher}{Association for
  Computing Machinery}, \bibinfo{address}{New York, NY, USA},
  \bibinfo{pages}{220–229}.
\newblock
\showISBNx{9781450361255}
\urldef\tempurl%
\url{https://doi.org/10.1145/3287560.3287596}
\showDOI{\tempurl}


\bibitem[\protect\citeauthoryear{Moreno, Aponte, Sridhara, Marcus, Pollock, and
  Vijay-Shanker}{Moreno et~al\mbox{.}}{2013}]%
        {moreno2013automatic}
\bibfield{author}{\bibinfo{person}{Laura Moreno}, \bibinfo{person}{Jairo
  Aponte}, \bibinfo{person}{Giriprasad Sridhara}, \bibinfo{person}{Andrian
  Marcus}, \bibinfo{person}{Lori Pollock}, {and} \bibinfo{person}{K.
  Vijay-Shanker}.} \bibinfo{year}{2013}\natexlab{}.
\newblock \showarticletitle{Automatic generation of natural language summaries
  for Java classes}. In \bibinfo{booktitle}{\emph{2013 21st International
  Conference on Program Comprehension (ICPC)}}. \bibinfo{publisher}{IEEE
  Press}, \bibinfo{pages}{23--32}.
\newblock
\urldef\tempurl%
\url{https://doi.org/10.1109/ICPC.2013.6613830}
\showDOI{\tempurl}


\bibitem[\protect\citeauthoryear{Muller, Lange, Wang, Piorkowski, Tsay, Liao,
  Dugan, and Erickson}{Muller et~al\mbox{.}}{2019}]%
        {muller2019data}
\bibfield{author}{\bibinfo{person}{Michael Muller}, \bibinfo{person}{Ingrid
  Lange}, \bibinfo{person}{Dakuo Wang}, \bibinfo{person}{David Piorkowski},
  \bibinfo{person}{Jason Tsay}, \bibinfo{person}{Q.~Vera Liao},
  \bibinfo{person}{Casey Dugan}, {and} \bibinfo{person}{Thomas Erickson}.}
  \bibinfo{year}{2019}\natexlab{}.
\newblock \showarticletitle{How Data Science Workers Work with Data: Discovery,
  Capture, Curation, Design, Creation}. In
  \bibinfo{booktitle}{\emph{Proceedings of the 2019 CHI Conference on Human
  Factors in Computing Systems}} (Glasgow, Scotland Uk)
  \emph{(\bibinfo{series}{CHI '19})}. \bibinfo{publisher}{Association for
  Computing Machinery}, \bibinfo{address}{New York, NY, USA},
  \bibinfo{pages}{1–15}.
\newblock
\showISBNx{9781450359702}
\urldef\tempurl%
\url{https://doi.org/10.1145/3290605.3300356}
\showDOI{\tempurl}


\bibitem[\protect\citeauthoryear{Nahar, Zhou, Lewis, and K\"{a}stner}{Nahar
  et~al\mbox{.}}{2022}]%
        {NZLK:ICSE22}
\bibfield{author}{\bibinfo{person}{Nadia Nahar}, \bibinfo{person}{Shurui Zhou},
  \bibinfo{person}{Grace Lewis}, {and} \bibinfo{person}{Christian
  K\"{a}stner}.} \bibinfo{year}{2022}\natexlab{}.
\newblock \showarticletitle{Collaboration Challenges in Building ML-Enabled
  Systems: Communication, Documentation, Engineering, and Process}. In
  \bibinfo{booktitle}{\emph{Proceedings of the 44th International Conference on
  Software Engineering}} (Pittsburgh, Pennsylvania)
  \emph{(\bibinfo{series}{ICSE '22})}. \bibinfo{publisher}{Association for
  Computing Machinery}, \bibinfo{address}{New York, NY, USA},
  \bibinfo{pages}{413–425}.
\newblock
\showISBNx{9781450392211}
\urldef\tempurl%
\url{https://doi.org/10.1145/3510003.3510209}
\showDOI{\tempurl}


\bibitem[\protect\citeauthoryear{Oracle}{Oracle}{[n.d.]}]%
        {JavaDoc}
\bibfield{author}{\bibinfo{person}{Oracle}.} \bibinfo{year}{[n.d.]}\natexlab{}.
\newblock \bibinfo{title}{How to Write Doc Comments for the Javadoc Tool}.
\newblock
  \bibinfo{howpublished}{\url{https://www.oracle.com/ca-en/technical-resources/articles/java/javadoc-tool.html}}.
\newblock


\bibitem[\protect\citeauthoryear{Patel, Fogarty, Landay, and Harrison}{Patel
  et~al\mbox{.}}{2008}]%
        {patel2008investigating}
\bibfield{author}{\bibinfo{person}{Kayur Patel}, \bibinfo{person}{James
  Fogarty}, \bibinfo{person}{James~A. Landay}, {and} \bibinfo{person}{Beverly
  Harrison}.} \bibinfo{year}{2008}\natexlab{}.
\newblock \showarticletitle{Investigating Statistical Machine Learning as a
  Tool for Software Development}. In \bibinfo{booktitle}{\emph{Proceedings of
  the SIGCHI Conference on Human Factors in Computing Systems}} (Florence,
  Italy) \emph{(\bibinfo{series}{CHI '08})}. \bibinfo{publisher}{Association
  for Computing Machinery}, \bibinfo{address}{New York, NY, USA},
  \bibinfo{pages}{667–676}.
\newblock
\showISBNx{9781605580111}
\urldef\tempurl%
\url{https://doi.org/10.1145/1357054.1357160}
\showDOI{\tempurl}


\bibitem[\protect\citeauthoryear{Pimentel, Murta, Braganholo, and
  Freire}{Pimentel et~al\mbox{.}}{2019}]%
        {pimentel2019large}
\bibfield{author}{\bibinfo{person}{Jo\~{a}o~Felipe Pimentel},
  \bibinfo{person}{Leonardo Murta}, \bibinfo{person}{Vanessa Braganholo}, {and}
  \bibinfo{person}{Juliana Freire}.} \bibinfo{year}{2019}\natexlab{}.
\newblock \showarticletitle{A Large-Scale Study about Quality and
  Reproducibility of Jupyter Notebooks}. In
  \bibinfo{booktitle}{\emph{Proceedings of the 16th International Conference on
  Mining Software Repositories}} (Montreal, Quebec, Canada)
  \emph{(\bibinfo{series}{MSR '19})}. \bibinfo{publisher}{IEEE Press},
  \bibinfo{pages}{507–517}.
\newblock
\urldef\tempurl%
\url{https://doi.org/10.1109/MSR.2019.00077}
\showDOI{\tempurl}


\bibitem[\protect\citeauthoryear{Prana, Treude, Thung, Atapattu, and Lo}{Prana
  et~al\mbox{.}}{2019}]%
        {prana2019categorizing}
\bibfield{author}{\bibinfo{person}{Gede Artha~Azriadi Prana},
  \bibinfo{person}{Christoph Treude}, \bibinfo{person}{Ferdian Thung},
  \bibinfo{person}{Thushari Atapattu}, {and} \bibinfo{person}{David Lo}.}
  \bibinfo{year}{2019}\natexlab{}.
\newblock \showarticletitle{Categorizing the content of github readme files}.
\newblock \bibinfo{journal}{\emph{Empirical Software Engineering}}
  \bibinfo{volume}{24}, \bibinfo{number}{3} (\bibinfo{year}{2019}),
  \bibinfo{pages}{1296--1327}.
\newblock


\bibitem[\protect\citeauthoryear{Psallidas, Zhu, Karlas, Interlandi, Floratou,
  Karanasos, Wu, Zhang, Krishnan, Curino, and Weimer}{Psallidas
  et~al\mbox{.}}{2019}]%
        {psallidas2019data}
\bibfield{author}{\bibinfo{person}{Fotis Psallidas}, \bibinfo{person}{Yiwen
  Zhu}, \bibinfo{person}{Bojan Karlas}, \bibinfo{person}{Matteo Interlandi},
  \bibinfo{person}{Avrilia Floratou}, \bibinfo{person}{Konstantinos Karanasos},
  \bibinfo{person}{Wentao Wu}, \bibinfo{person}{Ce Zhang},
  \bibinfo{person}{Subru Krishnan}, \bibinfo{person}{Carlo Curino}, {and}
  \bibinfo{person}{Markus Weimer}.} \bibinfo{year}{2019}\natexlab{}.
\newblock \showarticletitle{Data Science through the looking glass and what we
  found there}.
\newblock  (\bibinfo{year}{2019}).
\newblock
\urldef\tempurl%
\url{https://doi.org/10.48550/ARXIV.1912.09536}
\showDOI{\tempurl}
\showeprint{arXiv:1912.09536}


\bibitem[\protect\citeauthoryear{Raji, Smart, White, Mitchell, Gebru,
  Hutchinson, Smith-Loud, Theron, and Barnes}{Raji et~al\mbox{.}}{2020}]%
        {10.1145/3351095.3372873}
\bibfield{author}{\bibinfo{person}{Inioluwa~Deborah Raji},
  \bibinfo{person}{Andrew Smart}, \bibinfo{person}{Rebecca~N. White},
  \bibinfo{person}{Margaret Mitchell}, \bibinfo{person}{Timnit Gebru},
  \bibinfo{person}{Ben Hutchinson}, \bibinfo{person}{Jamila Smith-Loud},
  \bibinfo{person}{Daniel Theron}, {and} \bibinfo{person}{Parker Barnes}.}
  \bibinfo{year}{2020}\natexlab{}.
\newblock \showarticletitle{Closing the AI Accountability Gap: Defining an
  End-to-End Framework for Internal Algorithmic Auditing}. In
  \bibinfo{booktitle}{\emph{Proceedings of the 2020 Conference on Fairness,
  Accountability, and Transparency}} (Barcelona, Spain)
  \emph{(\bibinfo{series}{FAT* '20})}. \bibinfo{publisher}{Association for
  Computing Machinery}, \bibinfo{address}{New York, NY, USA},
  \bibinfo{pages}{33–44}.
\newblock
\showISBNx{9781450369367}
\urldef\tempurl%
\url{https://doi.org/10.1145/3351095.3372873}
\showDOI{\tempurl}


\bibitem[\protect\citeauthoryear{Richards, Piorkowski, Hind, Houde, and
  Mojsilović}{Richards et~al\mbox{.}}{2020}]%
        {richards2020factsheets}
\bibfield{author}{\bibinfo{person}{John Richards}, \bibinfo{person}{David
  Piorkowski}, \bibinfo{person}{Michael Hind}, \bibinfo{person}{Stephanie
  Houde}, {and} \bibinfo{person}{Aleksandra Mojsilović}.}
  \bibinfo{year}{2020}\natexlab{}.
\newblock \bibinfo{title}{A Methodology for Creating AI FactSheets}.
\newblock
\newblock
\urldef\tempurl%
\url{https://doi.org/10.48550/ARXIV.2006.13796}
\showDOI{\tempurl}
\showeprint{arXiv:2006.13796}


\bibitem[\protect\citeauthoryear{Robillard, Marcus, Treude, Bavota, Chaparro,
  Ernst, Gerosa, Godfrey, Lanza, Linares-Vásquez, Murphy, Moreno, Shepherd,
  and Wong}{Robillard et~al\mbox{.}}{2017}]%
        {robillard2017demand}
\bibfield{author}{\bibinfo{person}{Martin~P. Robillard},
  \bibinfo{person}{Andrian Marcus}, \bibinfo{person}{Christoph Treude},
  \bibinfo{person}{Gabriele Bavota}, \bibinfo{person}{Oscar Chaparro},
  \bibinfo{person}{Neil Ernst}, \bibinfo{person}{Marco~Aurélio Gerosa},
  \bibinfo{person}{Michael Godfrey}, \bibinfo{person}{Michele Lanza},
  \bibinfo{person}{Mario Linares-Vásquez}, \bibinfo{person}{Gail~C. Murphy},
  \bibinfo{person}{Laura Moreno}, \bibinfo{person}{David Shepherd}, {and}
  \bibinfo{person}{Edmund Wong}.} \bibinfo{year}{2017}\natexlab{}.
\newblock \showarticletitle{On-demand Developer Documentation}. In
  \bibinfo{booktitle}{\emph{2017 IEEE International Conference on Software
  Maintenance and Evolution (ICSME)}}. \bibinfo{publisher}{IEEE Press},
  \bibinfo{pages}{479--483}.
\newblock
\urldef\tempurl%
\url{https://doi.org/10.1109/ICSME.2017.17}
\showDOI{\tempurl}


\bibitem[\protect\citeauthoryear{Rule, Tabard, and Hollan}{Rule
  et~al\mbox{.}}{2018}]%
        {github2017dataset}
\bibfield{author}{\bibinfo{person}{Adam Rule}, \bibinfo{person}{Aur\'{e}lien
  Tabard}, {and} \bibinfo{person}{James~D. Hollan}.}
  \bibinfo{year}{2018}\natexlab{}.
\newblock \showarticletitle{Exploration and Explanation in Computational
  Notebooks}. In \bibinfo{booktitle}{\emph{Proceedings of the 2018 CHI
  Conference on Human Factors in Computing Systems}} (Montreal QC, Canada)
  \emph{(\bibinfo{series}{CHI '18})}. \bibinfo{publisher}{Association for
  Computing Machinery}, \bibinfo{address}{New York, NY, USA},
  \bibinfo{pages}{1–12}.
\newblock
\showISBNx{9781450356206}
\urldef\tempurl%
\url{https://doi.org/10.1145/3173574.3173606}
\showDOI{\tempurl}


\bibitem[\protect\citeauthoryear{Sculley, Holt, Golovin, Davydov, Phillips,
  Ebner, Chaudhary, Young, Crespo, and Dennison}{Sculley et~al\mbox{.}}{2015}]%
        {sculley2015hidden}
\bibfield{author}{\bibinfo{person}{D. Sculley}, \bibinfo{person}{Gary Holt},
  \bibinfo{person}{Daniel Golovin}, \bibinfo{person}{Eugene Davydov},
  \bibinfo{person}{Todd Phillips}, \bibinfo{person}{Dietmar Ebner},
  \bibinfo{person}{Vinay Chaudhary}, \bibinfo{person}{Michael Young},
  \bibinfo{person}{Jean-Fran\c{c}ois Crespo}, {and} \bibinfo{person}{Dan
  Dennison}.} \bibinfo{year}{2015}\natexlab{}.
\newblock \showarticletitle{Hidden Technical Debt in Machine Learning Systems}.
  In \bibinfo{booktitle}{\emph{Advances in Neural Information Processing
  Systems}}, \bibfield{editor}{\bibinfo{person}{C.~Cortes},
  \bibinfo{person}{N.~Lawrence}, \bibinfo{person}{D.~Lee},
  \bibinfo{person}{M.~Sugiyama}, {and} \bibinfo{person}{R.~Garnett}} (Eds.),
  Vol.~\bibinfo{volume}{28}. \bibinfo{publisher}{Curran Associates, Inc.}
\newblock
\urldef\tempurl%
\url{https://proceedings.neurips.cc/paper/2015/file/86df7dcfd896fcaf2674f757a2463eba-Paper.pdf}
\showURL{%
\tempurl}


\bibitem[\protect\citeauthoryear{Siebert, Joeckel, Heidrich, Trendowicz,
  Nakamichi, Ohashi, Namba, Yamamoto, and Aoyama}{Siebert
  et~al\mbox{.}}{2022}]%
        {siebert2022construction}
\bibfield{author}{\bibinfo{person}{Julien Siebert}, \bibinfo{person}{Lisa
  Joeckel}, \bibinfo{person}{Jens Heidrich}, \bibinfo{person}{Adam Trendowicz},
  \bibinfo{person}{Koji Nakamichi}, \bibinfo{person}{Kyoko Ohashi},
  \bibinfo{person}{Isao Namba}, \bibinfo{person}{Rieko Yamamoto}, {and}
  \bibinfo{person}{Mikio Aoyama}.} \bibinfo{year}{2022}\natexlab{}.
\newblock \showarticletitle{Construction of a quality model for machine
  learning systems}.
\newblock \bibinfo{journal}{\emph{Software Quality Journal}}
  \bibinfo{volume}{30}, \bibinfo{number}{2} (\bibinfo{year}{2022}),
  \bibinfo{pages}{307--335}.
\newblock


\bibitem[\protect\citeauthoryear{Sommerville}{Sommerville}{2016}]%
        {Ian2016SE}
\bibfield{author}{\bibinfo{person}{Ian Sommerville}.}
  \bibinfo{year}{2016}\natexlab{}.
\newblock \bibinfo{booktitle}{\emph{Software Engineering, 10th edition}}.
\newblock \bibinfo{publisher}{Pearson}.
\newblock


\bibitem[\protect\citeauthoryear{Star and Griesemer}{Star and
  Griesemer}{1989}]%
        {star1989institutional}
\bibfield{author}{\bibinfo{person}{Susan~Leigh Star} {and}
  \bibinfo{person}{James~R Griesemer}.} \bibinfo{year}{1989}\natexlab{}.
\newblock \showarticletitle{Institutional ecology,translations' and boundary
  objects: Amateurs and professionals in Berkeley's Museum of Vertebrate
  Zoology, 1907-39}.
\newblock \bibinfo{journal}{\emph{Social studies of science}}
  \bibinfo{volume}{19}, \bibinfo{number}{3} (\bibinfo{year}{1989}),
  \bibinfo{pages}{387--420}.
\newblock


\bibitem[\protect\citeauthoryear{Stolee, Elbaum, and Dwyer}{Stolee
  et~al\mbox{.}}{2016}]%
        {STOLEE201635}
\bibfield{author}{\bibinfo{person}{Kathryn~T. Stolee},
  \bibinfo{person}{Sebastian Elbaum}, {and} \bibinfo{person}{Matthew~B.
  Dwyer}.} \bibinfo{year}{2016}\natexlab{}.
\newblock \showarticletitle{Code search with input/output queries:
  Generalizing, ranking, and assessment}.
\newblock \bibinfo{journal}{\emph{Journal of Systems and Software}}
  \bibinfo{volume}{116} (\bibinfo{year}{2016}), \bibinfo{pages}{35--48}.
\newblock
\showISSN{0164-1212}
\urldef\tempurl%
\url{https://doi.org/10.1016/j.jss.2015.04.081}
\showDOI{\tempurl}


\bibitem[\protect\citeauthoryear{Stylos and Myers}{Stylos and Myers}{2006}]%
        {stylos2006mica}
\bibfield{author}{\bibinfo{person}{J. Stylos} {and} \bibinfo{person}{B.A.
  Myers}.} \bibinfo{year}{2006}\natexlab{}.
\newblock \showarticletitle{Mica: A Web-Search Tool for Finding API Components
  and Examples}. In \bibinfo{booktitle}{\emph{Visual Languages and
  Human-Centric Computing (VL/HCC'06)}}. \bibinfo{publisher}{IEEE Press},
  \bibinfo{pages}{195--202}.
\newblock
\urldef\tempurl%
\url{https://doi.org/10.1109/VLHCC.2006.32}
\showDOI{\tempurl}


\bibitem[\protect\citeauthoryear{Treude and Robillard}{Treude and
  Robillard}{2016}]%
        {treude2016augmenting}
\bibfield{author}{\bibinfo{person}{Christoph Treude} {and}
  \bibinfo{person}{Martin~P. Robillard}.} \bibinfo{year}{2016}\natexlab{}.
\newblock \showarticletitle{Augmenting API Documentation with Insights from
  Stack Overflow}. In \bibinfo{booktitle}{\emph{Proceedings of the 38th
  International Conference on Software Engineering}} (Austin, Texas)
  \emph{(\bibinfo{series}{ICSE '16})}. \bibinfo{publisher}{Association for
  Computing Machinery}, \bibinfo{address}{New York, NY, USA},
  \bibinfo{pages}{392–403}.
\newblock
\showISBNx{9781450339001}
\urldef\tempurl%
\url{https://doi.org/10.1145/2884781.2884800}
\showDOI{\tempurl}


\bibitem[\protect\citeauthoryear{Uddin and Robillard}{Uddin and
  Robillard}{2015}]%
        {7140676}
\bibfield{author}{\bibinfo{person}{Gias Uddin} {and} \bibinfo{person}{Martin~P.
  Robillard}.} \bibinfo{year}{2015}\natexlab{}.
\newblock \showarticletitle{How API Documentation Fails}.
\newblock \bibinfo{journal}{\emph{IEEE Software}} \bibinfo{volume}{32},
  \bibinfo{number}{4} (\bibinfo{year}{2015}), \bibinfo{pages}{68--75}.
\newblock
\urldef\tempurl%
\url{https://doi.org/10.1109/MS.2014.80}
\showDOI{\tempurl}


\bibitem[\protect\citeauthoryear{Vaismoradi, Turunen, and Bondas}{Vaismoradi
  et~al\mbox{.}}{2013}]%
        {Vaismoradi2013}
\bibfield{author}{\bibinfo{person}{Mojtaba Vaismoradi},
  \bibinfo{person}{Hannele Turunen}, {and} \bibinfo{person}{Terese Bondas}.}
  \bibinfo{year}{2013}\natexlab{}.
\newblock \showarticletitle{Content analysis and thematic analysis:
  Implications for conducting a qualitative descriptive study}.
\newblock \bibinfo{journal}{\emph{Nursing \& Health Sciences}}
  \bibinfo{volume}{15}, \bibinfo{number}{3} (\bibinfo{year}{2013}),
  \bibinfo{pages}{398--405}.
\newblock
\urldef\tempurl%
\url{https://doi.org/10.1111/nhs.12048}
\showDOI{\tempurl}
\showeprint{https://onlinelibrary.wiley.com/doi/pdf/10.1111/nhs.12048}


\bibitem[\protect\citeauthoryear{Wang, Epperson, DeLine, and Drucker}{Wang
  et~al\mbox{.}}{2022a}]%
        {wang2022diff}
\bibfield{author}{\bibinfo{person}{April~Yi Wang}, \bibinfo{person}{Will
  Epperson}, \bibinfo{person}{Robert~A DeLine}, {and}
  \bibinfo{person}{Steven~M. Drucker}.} \bibinfo{year}{2022}\natexlab{a}.
\newblock \showarticletitle{Diff in the Loop: Supporting Data Comparison in
  Exploratory Data Analysis}. In \bibinfo{booktitle}{\emph{Proceedings of the
  2022 CHI Conference on Human Factors in Computing Systems}} (New Orleans, LA,
  USA) \emph{(\bibinfo{series}{CHI '22})}. \bibinfo{publisher}{Association for
  Computing Machinery}, \bibinfo{address}{New York, NY, USA}, Article
  \bibinfo{articleno}{97}, \bibinfo{numpages}{10}~pages.
\newblock
\showISBNx{9781450391573}
\urldef\tempurl%
\url{https://doi.org/10.1145/3491102.3502123}
\showDOI{\tempurl}


\bibitem[\protect\citeauthoryear{Wang, Wang, Drozdal, Muller, Park, Weisz, Liu,
  Wu, and Dugan}{Wang et~al\mbox{.}}{2022b}]%
        {wang2021documentation}
\bibfield{author}{\bibinfo{person}{April~Yi Wang}, \bibinfo{person}{Dakuo
  Wang}, \bibinfo{person}{Jaimie Drozdal}, \bibinfo{person}{Michael Muller},
  \bibinfo{person}{Soya Park}, \bibinfo{person}{Justin~D. Weisz},
  \bibinfo{person}{Xuye Liu}, \bibinfo{person}{Lingfei Wu}, {and}
  \bibinfo{person}{Casey Dugan}.} \bibinfo{year}{2022}\natexlab{b}.
\newblock \showarticletitle{Documentation Matters: Human-Centered AI System to
  Assist Data Science Code Documentation in Computational Notebooks}.
\newblock \bibinfo{journal}{\emph{ACM Trans. Comput.-Hum. Interact.}}
  \bibinfo{volume}{29}, \bibinfo{number}{2}, Article \bibinfo{articleno}{17}
  (\bibinfo{date}{jan} \bibinfo{year}{2022}), \bibinfo{numpages}{33}~pages.
\newblock
\showISSN{1073-0516}
\urldef\tempurl%
\url{https://doi.org/10.1145/3489465}
\showDOI{\tempurl}


\bibitem[\protect\citeauthoryear{Weinman, Drucker, Barik, and DeLine}{Weinman
  et~al\mbox{.}}{2021}]%
        {nathaniel2022fork}
\bibfield{author}{\bibinfo{person}{Nathaniel Weinman},
  \bibinfo{person}{Steven~M. Drucker}, \bibinfo{person}{Titus Barik}, {and}
  \bibinfo{person}{Robert DeLine}.} \bibinfo{year}{2021}\natexlab{}.
\newblock \showarticletitle{Fork It: Supporting Stateful Alternatives in
  Computational Notebooks}. In \bibinfo{booktitle}{\emph{Proceedings of the
  2021 CHI Conference on Human Factors in Computing Systems}} (Yokohama, Japan)
  \emph{(\bibinfo{series}{CHI '21})}. \bibinfo{publisher}{Association for
  Computing Machinery}, \bibinfo{address}{New York, NY, USA}, Article
  \bibinfo{articleno}{307}, \bibinfo{numpages}{12}~pages.
\newblock
\showISBNx{9781450380966}
\urldef\tempurl%
\url{https://doi.org/10.1145/3411764.3445527}
\showDOI{\tempurl}


\bibitem[\protect\citeauthoryear{Yang, Zhou, Guo, and K\"{a}stner}{Yang
  et~al\mbox{.}}{2022}]%
        {yang2021subtle}
\bibfield{author}{\bibinfo{person}{Chenyang Yang}, \bibinfo{person}{Shurui
  Zhou}, \bibinfo{person}{Jin L.~C. Guo}, {and} \bibinfo{person}{Christian
  K\"{a}stner}.} \bibinfo{year}{2022}\natexlab{}.
\newblock \showarticletitle{Subtle Bugs Everywhere: Generating Documentation
  for Data Wrangling Code}. In \bibinfo{booktitle}{\emph{Proceedings of the
  36th IEEE/ACM International Conference on Automated Software Engineering}}
  (Melbourne, Australia) \emph{(\bibinfo{series}{ASE '21})}.
  \bibinfo{publisher}{IEEE Press}, \bibinfo{pages}{304–316}.
\newblock
\showISBNx{9781665403375}
\urldef\tempurl%
\url{https://doi.org/10.1109/ASE51524.2021.9678520}
\showDOI{\tempurl}


\bibitem[\protect\citeauthoryear{Zhang, Muller, and Wang}{Zhang
  et~al\mbox{.}}{2020}]%
        {zhang2020data}
\bibfield{author}{\bibinfo{person}{Amy~X Zhang}, \bibinfo{person}{Michael
  Muller}, {and} \bibinfo{person}{Dakuo Wang}.}
  \bibinfo{year}{2020}\natexlab{}.
\newblock \showarticletitle{How do data science workers collaborate? roles,
  workflows, and tools}.
\newblock \bibinfo{journal}{\emph{Proceedings of the ACM on Human-Computer
  Interaction}} \bibinfo{volume}{4}, \bibinfo{number}{CSCW1}
  (\bibinfo{year}{2020}), \bibinfo{pages}{1--23}.
\newblock


\end{thebibliography}

\end{document}